\documentclass[lettersize,journal]{IEEEtran}
\usepackage{amsmath,amsfonts}
\usepackage{amsmath,amssymb,amsfonts}
\usepackage{algorithmicx}
\usepackage{algorithm}
\usepackage{array}
\usepackage{algpseudocode}
\usepackage[caption=false,font=normalsize,labelfont=sf,textfont=sf]{subfig}
\usepackage{textcomp}
\usepackage{stfloats}
\usepackage{framed,multirow,color}
\usepackage{url}
\usepackage{verbatim}
\usepackage{threeparttable}
\usepackage{graphicx}
\def\BibTeX{{\rm B\kern-.05em{\sc i\kern-.025em b}\kern-.08em
    T\kern-.1667em\lower.7ex\hbox{E}\kern-.125emX}}
\usepackage{balance}

\begin{document}
\title{Enhancing and Adapting in the Clinic: Source-free Unsupervised Domain Adaptation for Medical Image Enhancement}

\author{
Heng Li, Ziqin Lin, Zhongxi Qiu, Zinan Li, Ke Niu, Na Guo, Huazhu Fu, Yan Hu, Jiang Liu
\thanks{This work was supported in part by the National Natural Science Foundation of China (82272086, 82102189), Guangdong Provincial Department of Education (SJZLGC202202), and Shenzhen Natural Science Fund (20200925174052004).
(Corresponding Author: Jiang Liu, liuj@sustech.edu.cn)
}
\thanks{H. Li, Z. Qiu, Y. Hu, and J. Liu are with the Research Institute of Trustworthy Autonomous Systems, Southern University of Science and Technology, Shenzhen, China; 
Z. Lin, Z. Li, and J. Liu are with the Department of Computer Science and Engineering, Southern University of Science and Technology, Shenzhen, China;
H. Fu is with the Institute of High Performance Computing, Agency for Science, Technology and Research, Singapore;
K. Niu is with Computer School, Beijing Information Science and Technology University, Beijing, China;
N. Guo is with the School of Computer and Communication Engineering, University of Science and Technology Beijing, Beijing, China.}
}

\maketitle

\begin{abstract}
Medical imaging provides many valuable clues involving anatomical structure and pathological characteristics.
However, image degradation is a common issue in clinical practice, which can adversely impact the observation and diagnosis by physicians and algorithms.
Although extensive enhancement models have been developed, these models require a well pre-training before deployment, while failing to take advantage of the potential value of inference data after deployment.
In this paper, we raise an algorithm for source-free unsupervised domain adaptive medical image enhancement (SAME), which adapts and optimizes enhancement models using test data in the inference phase.
A structure-preserving enhancement network is first constructed to learn a robust source model from synthesized training data.
Then a teacher-student model is initialized with the source model and conducts source-free unsupervised domain adaptation (SFUDA) by knowledge distillation with the test data.
Additionally, a pseudo-label picker is developed to boost the knowledge distillation of enhancement tasks.
Experiments were implemented on ten datasets from three medical image modalities to validate the advantage of the proposed algorithm, and setting analysis and ablation studies were also carried out to interpret the effectiveness of SAME.
The remarkable enhancement performance and benefits for downstream tasks demonstrate the potential and generalizability of SAME.
The code is available at https://github.com/liamheng/Annotation-free-Medical-Image-Enhancement.
\end{abstract}

\begin{IEEEkeywords}
Medical image enhancement, source-free unsupervised domain adaptation, knowledge distillation, pseudo-label selection.
\end{IEEEkeywords}

\section{Introduction}
\begin{figure}[t]
\begin{centering}
\includegraphics[width=1\linewidth]{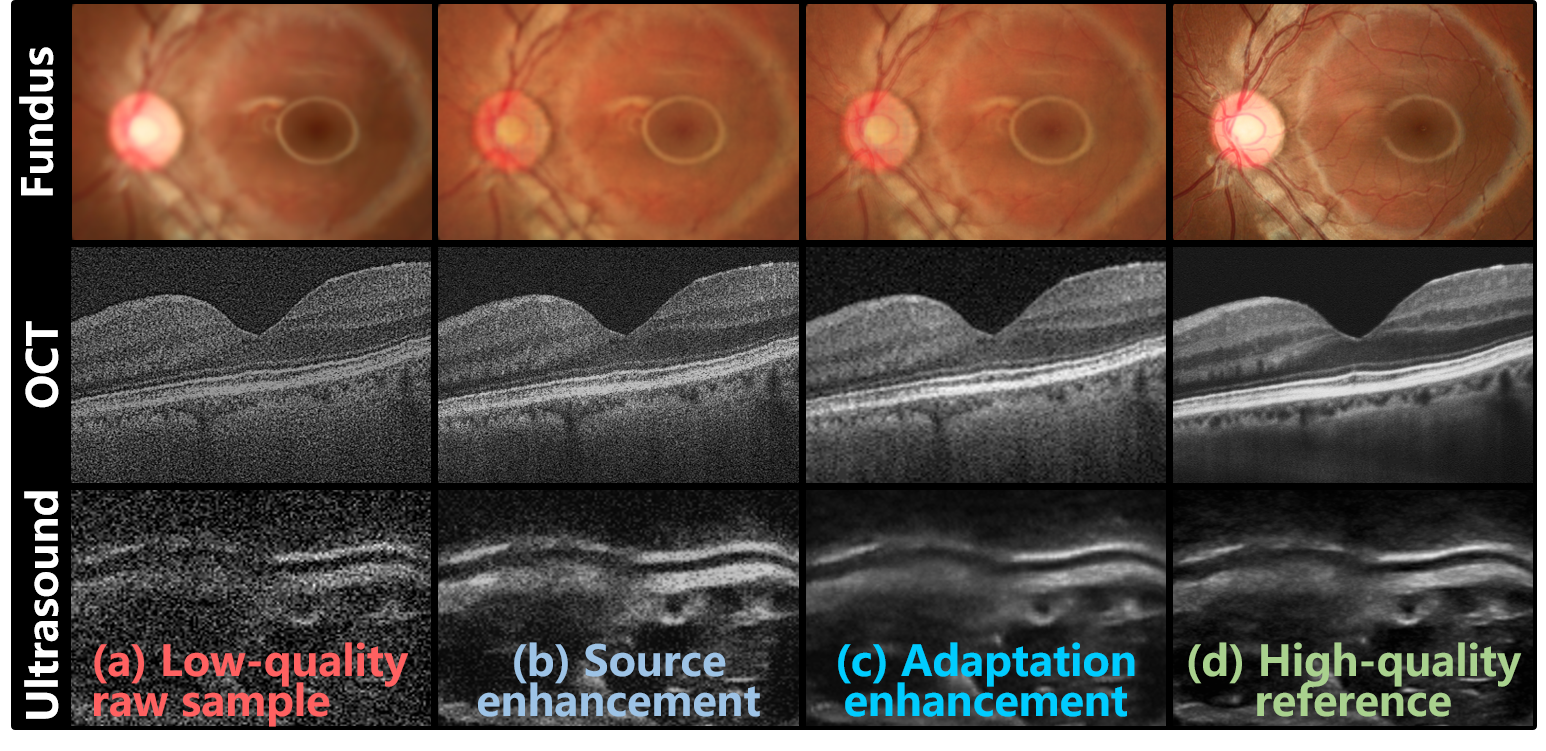}
\par\end{centering}
\caption{Examples of medical image enhancement. 
(a) Low-quality samples of fundus photography, OCT, and ultrasound.
(b) Samples enhanced by the source model and (c) further improved by SFUDA.
(d) High-quality references.
}
\label{fig:introduction}
\end{figure}

The advancement of modern medical imaging technology has provided a wealth of valuable clues regarding anatomical structures and pathological characteristics for disease diagnosis. 
Based on high-quality medical images, emerging deep learning algorithms have demonstrated significant potential in medical image analysis and disease diagnosis, achieving a diagnostic performance comparable to that of human medical professionals~\cite{li2021applications}.
Regrettably, medical imaging in clinical settings is vulnerable to quality degradation (Fig.~\ref{fig:introduction} (a)) caused by environmental factors, inappropriate operation, and patient status~\cite{conzelmann2022comparison}, leading to uncertainties in observations and diagnoses~\cite{liu2022understanding}. 
For instance, an investigation of 135,867 fundus images in the UK Biobank revealed that only 71.5\% of the samples were of sufficient quality to conduct vessel morphometry~\cite{welikala2017automated}.
In such cases, patients may need to be re-examined to obtain qualified imaging data, causing unnecessary costs, secondary radiation exposure, and time delays.

Enhancing the legibility of low-quality medical imaging holds great potential in improving clinical observation and diagnosis, avoiding the shortcomings of re-examination.
Consequently, significant efforts have been devoted to enhancing medical images over the years.
Historically, researchers have mined statistical prior knowledge to develop hand-crafted methods for medical image enhancement~\cite{mitra2018enhancement}. 
With recent advances in deep learning, powerful image embedding capabilities have enabled more convenient and efficient enhancement of medical images.
Despite the advantages of deep learning, abundant training data are necessary to optimize deep networks.
Furthermore, enhancement algorithms typically rely on high-low quality paired data, which creates a major challenge in data collection~\cite{RFormer}.
To address this issue, more recent studies have introduced enhancement algorithms based on unpaired~\cite{ma2021structure,cheng2021secret} and synthetic data~\cite{li2022annotation,shen2020modeling}. 
However, these methods have their own respective limitations, such as compromised image structure preservation with unpaired data or performance drops in real-world scenarios with synthesized data.
As a result, existing enhancement methods either suffer from data collection issues or fail to guarantee consistent performance in clinical settings~\cite{li2023generic}.
Moreover, these methods rely on well pre-training before deployment and lack the flexibility to learn and adapt from inference data after deployment.

While current methods have successfully improved the quality of medical images, clinical scenarios still pose several challenges that need further exploration.
i) Unpaired data-based algorithms overcome the need for repeatedly acquiring high-low quality image pairs. 
However, the distribution homogeneity of training data and clinical data cannot be guaranteed, and unpaired training data may not be conducive to preserving fine structures in enhancement models.
ii) Visiting target test data during model training is a crucial technique for bridging the domain shifts between synthetic data and real data. 
However, obtaining target data in advance for centralized training may raise concerns about data collection and privacy in clinical scenarios.
iii) Existing enhancement algorithms rely on well-pre-trained models to promote clinical observation and diagnosis, due to privacy concerns and implementation feasibility. 
However, this strategy may not fully overcome the clinical data distribution shifts caused by various scanners, protocols, and patient demographics.
iv) Even when enhancement models are exposed to target data in clinical applications,  the models often lack the ability to leverage the data for further learning and performance optimization.

To alleviate the above challenges, this paper introduces \textbf{S}ource-free unsupervised domain \textbf{A}daptive \textbf{M}edical image \textbf{E}nhancement (SAME), which adapts and optimizes enhancement models using test data in the inference phase.
Specifically, to initialize a robust enhancement model with structure preservation (Fig.~\ref{fig:introduction} (b)), the source model is trained with a source domain synthesized from public datasets along with segmentation masks.
Subsequently, in the target domain of test data, a teacher-student model is employed to perform knowledge distillation to further optime the enhancement model in the inference phase (Fig.~\ref{fig:introduction} (c)).
In addition, a comprehensive picker has been developed, which includes an image quality assessor and an irregular structure detector, to select appropriate pseudo-labels for knowledge distillation.
Experiments on three medical image modalities (i.e., fundus photography, OCT, and ultrasound) have been conducted to demonstrate the enhancement performance of SAME.
Our main contributions are as follows:
\begin{itemize}
  \item [1)] A medical image enhancement algorithm termed SAME is developed, which introduces an SFUDA paradigm to optimize enhancement models in the inference phase without concerns about data collection and privacy.
  \item [2)] Initialized by the source model from synthetic training data, a teacher-student model is designed to achieve SFUDA through knowledge distillation. 
  \item [3)] Based on image quality and structural regularity, a customized picker that selects pseudo-labels for knowledge distillation is designed to boost the SFUDA in enhancement tasks.
  \item [4)] Various experiments and comparisons with diverse state-of-the-art (SOTA) enhancement algorithms are presented on three medical image modalities, and the benefits of SAME are demonstrated by the superior performance.
\end{itemize}


\section{Related Work}

\subsection{Medical image enhancement}
Owing to the ability to enhance the quality of clinical imaging examinations in a cost-effective and efficient manner, image enhancement has been a longstanding area of research in the medical imaging community.
In pioneering studies, statistical analysis has been extensively used to discover prior knowledge and develop hand-crafted enhancement algorithms. 
To improve image readability, contrast limited adaptive histogram equalization (CLAHE)~\cite{zuiderveld1994contrast} was designed to expand the dynamic ranges of images.
And then CLAHE has been applied to enhance fundus images~\cite{mitra2018enhancement}.
Inspired by guided image filtering (GIF)~\cite{he2012guided}, Cheng et al.~\cite{cheng2018structure} developed structure-preserving guided retinal image filtering (SGRIF) to restore cataract-affected fundus images as well as preserve fine structures.
However, traditional prior-based methods may not be suitable for solving multiple low-quality cases, as they are aimed at specific degradation types.

With the advances in deep learning, its powerful image embedding capability enables more convenient and efficient enhancement of medical images.
However, new challenges in data collection~\cite{RFormer} are also introduced in data requirements for supervised training deep learning neural networks.
Therefore, unpaired data and synthetic data are recently explored to optimize enhancement deep networks.

\subsubsection{Unpaired data-based medical image enhancement}
Compared with paired medical images, unpaired ones can be obtained more efficiently in clinical settings.
Based on the unpaired image translation via CycleGAN~\cite{zhu2017unpaired}, HDcycleGAN~\cite{manakov2019noise} and StillGAN~\cite{ma2021structure} were developed to enhance medical images by bridging the gap between low-quality and high-quality domains.
Another unpaired image translation network, known as contrastive unpaired translation (CUT)~\cite{park2020contrastive}, has been proposed using contrastive learning, which has inspired the development of medical image enhancement methods like I-SECRET~\cite{cheng2021secret}.
However, learning with unpaired data may not effectively preserve fine structures and may overlook distribution shifts, leading to suboptimal enhancement performance.

\subsubsection{Synthetic data-based medical image enhancement}
Alternatively, synthesizing high-low quality paired data has been frequently introduced to conduct supervised learning for medical image enhancement.
CofeNet~\cite{shen2020modeling} and ArcNet~\cite{li2022annotation} constructed several degradation models of fundus photographs by analyzing the imaging interference and imaging optical path of cataract patients, and subsequently proposed algorithms for fundus image enhancement.
Through fusing the surrounding b-scans, ODDM~\cite{hu2022unsupervised} synthesizes high-quality references to train a diffusion model for OCT denoising.
Notably, the performance of models trained on synthetic data may be impacted by shifts between synthetic and real-world domains.
Consequently, domain adaptation has been introduced in ArcNet~\cite{li2022annotation} and MAGE-Net~\cite{guo2023bridging}, which incorporate test data in the training phase to generalize models from synthetic to real-world data.
However, visiting test data for model training is often impractical, as it poses challenges in terms of data collection and privacy protection. 
Moreover, to further alleviate data dependency,  SCR-Net~\cite{li2022structure} and PCE-Net~\cite{liu2022degradation} constrain the representation consistency across various degradation views from identical images to impose the generalizability of enhancement models across synthetic domains.

Despite the advantages of current algorithms, they typically adopt a pipeline that involves well-pre-trained models to enhance medical images.
Due to the high costs and privacy concerns associated with data collection, most of the clinical data cannot be used by this pipeline, resulting in a scarcity of available training data.
Moreover, once deployed in clinics, models that rely on this pipeline may not be able to adapt to new data and optimize their performance accordingly.

\subsection{Source-free unsupervised domain adaptation}
In deep learning, a common cause of performance degradation is domain shifts, which refer to a variance in distribution between the source and target domains~\cite{zhou2022domain,li2023frequency}.
Aiming to this issue, domain adaptation~\cite{guan2021domain} has been proposed to perform knowledge transfer by reducing inter-domain distribution discrepancy.
However, as typical domain adaptation highly depends on the accessibility of both source and target data, practical limitations are inevitable, such as privacy concerns, data storage and transmission costs, and computation burden. 

To overcome these limitations, source-free unsupervised domain adaptation (SFUDA)~\cite{liang2023comprehensive} transfers a pre-trained source model to the unlabeled target domain without requiring any source data.
The goal of SFUDA is to improve target inference by learning a target model based on the pre-trained source model and unlabeled target data~\cite{fang2022source}. 
{
Prominent SFUDA paradigms, such as entropy minimization~\cite{liang2020we} and knowledge distillation~\cite{chen2022self}, have been developed in emerging studies to achieve adaptation solely using target data.
These paradigms have primarily focused on tasks like semantic segmentation~\cite{shin2022mm}, image classification~\cite{zhao2023source}, and object detection~\cite{liu2022source}, demonstrating promising progress.
Furthermore, recent developments in SFUDA have given rise to test-time adaptation (TTA) algorithms like TENT~\cite{wang2020tent} and MEMO~\cite{zhang2022memo}, which exhibit the potential to generalize well across diverse target domains without relying on target training data.
However, challenges arise when applying the above paradigms to image enhancement, since they typically rely on the category labels in the above tasks. 
Therefore, additional exploration is necessary to implement SFUDA for image enhancement purposes.
While a recent study has introduced domain representation normalization (DRN)~\cite{yu2022source} to apply SFUDA for natural image dehazing, there is a scarcity of reporting on the SFUDA paradigm for enhancing medical images. 
As a result, further efforts and investigations are required in this particular area.
}

Consequently, we attempt to develop an SFUDA  paradigm in medical image enhancement to mitigate the challenges in data collection and privacy protection when implementing domain adaptation in clinics.
Additionally, SFUDA endows to further optimize the enhancement model during the inference phase, mining the full potential of clinical data.

\section{Method}
\begin{figure*}[htbp]
\begin{centering}
\includegraphics[width=0.9\linewidth]{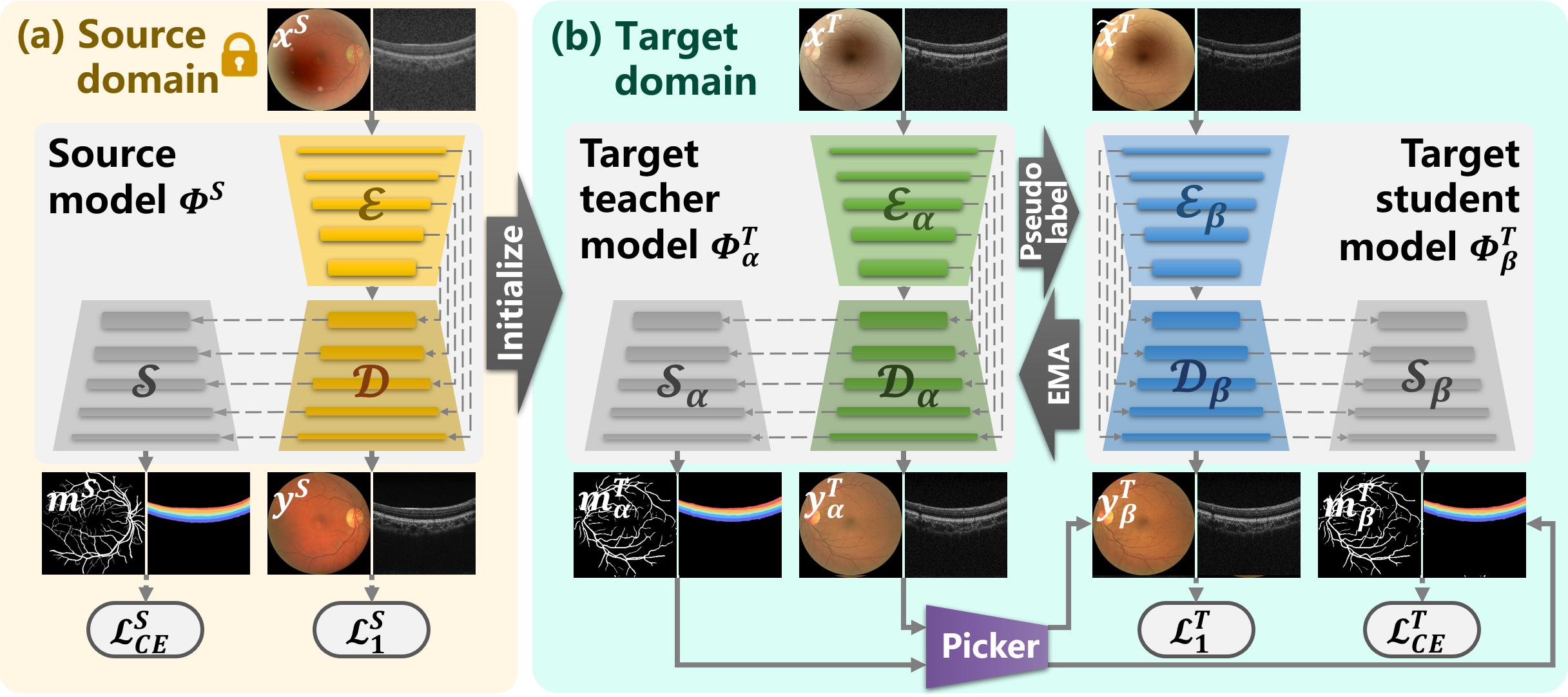}
\par\end{centering}
\caption{
SAME introduces SFUDA to optimize medical image enhancement models in clinical inference. 
(a) A source enhancement model $\Phi ^S$ is learned in the source domain of synthetic training data.
The backbone of $\Phi^S$ consists of a skip-connected encoder $\mathcal{E}$ and decoder $\mathcal{D}$, which enhance $x^S$ as $y^S$. And a segmenter $\mathcal{S}$ is connected to $\mathcal{D}$ to constrain the structural components $m^S$ of $y^S$.
(b) SFUDA is accomplished in the target domain through knowledge distillation, using the teacher-student model ($\Phi ^T_\alpha$ and $\Phi ^T_\beta$) initialized by $\Phi^S$. 
Specifically, an enhancement pseudo-label picker (more details in Fig.~\ref{fig:label}) is customized to boost the process of knowledge distillation, 
{and $\Phi ^T_\alpha$ is updated using the exponential moving average (EMA) from the weights of $\Phi ^T_\beta$.}
}
\label{fig:workflow}
\end{figure*}

Given the clinical challenges involved in enhancing medical images, SAME has been developed to overcome data bottlenecks and privacy concerns that arise when leveraging target data to optimize enhancement models. 
As demonstrated in Fig.~\ref{fig:workflow}, SAME achieves this by introducing SFUDA into medical image enhancement, where model training is only based on source data while target data are used to fine-tune the model during the inference phase.

\subsection{Preliminary and problem definition}
As a result of the difficulties in data collection, training data are frequently synthesized to develop medical image enhancement algorithms.
Denote high-quality medical image samples as $Y^S$, the source domain $\mathbb{D} ^S=\left \{ X^S,Y^S \right \}$ is composed of synthesized training data, where $X^S$ refers to the low-quality samples generated from $Y^S$ with degradation models.
Then a source model $\Phi ^S$ is well-trained based on the source domain to estimate the joint distribution $P^S_{XY}$ on $X^S\times Y^S$.
On the other hand, the target domain $\mathbb{D}^T=\left \{ X^T \right \}$ only contains clinical low-quality samples $X^T$ to be enhanced.
Unfortunately, due to the domain shifts between $\mathbb{D} ^S$ and $\mathbb{D} ^T$, the joint distribution $P^T_{XY} \ne P^S_{XY}$, where $P^T_{XY}$ is the joint distribution on $X^T\times Y^T$, $Y^T$ denotes the corresponding ideal high-quality samples.
Thus $\Phi ^S$ barely presents desirable performance on $\mathbb{D}^T$.

Though simultaneously visiting $\left \{ X^S,Y^S \right \}$ and $\left \{X^T\right \}$ allows DA to generalize $\Phi ^S$ from $\mathbb{D}^S$ to $\mathbb{D}^T$, the raised feasibility and privacy concerns prevent implementing DA in the clinic.
SAME resorts to SFUDA to address the above limitations, where $\left \{ X^S,Y^S \right \}$ is inaccessible but $\Phi ^S$ is available to conduct domain adaptation.

{To train a structure-preserving enhancement model $\Phi ^S$, public high-quality image samples along with segmentation masks are employed in SAME to synthesize the source domain  $\left \{ X^S, Y^S, M^S \right \}$, where $M^S$ denotes the segmentation masks.}
Then in the target domain, SAME initializes a teacher model $\Phi ^T_\alpha$ using the parameters of $\Phi ^S$.
This teacher model performs inference on the unannotated target data $X^T$, and the resulting outcomes $\left \{ Y^T_\alpha, M^T_\alpha \right \}$ are leveraged to pick pseudo-labels for fine-tuning a student model $\Phi ^T_\beta$.
Accordingly, SAME employs knowledge distillation by using the teacher-student model to achieve SFUDA.

\subsection{Structure-preserving source model} 
The organ structures in medical images often contain important diagnosis clues related to diseases. Therefore, authentically preserving organ structures in raw images is a fundamental prerequisite for effective medical image enhancement.

Inspired by the degradation models in~\cite{shen2020modeling,li2022annotation}, we synthesize low-quality images from public high-quality samples to construct training data for medical image enhancement.
Moreover, to preserve the organ structures of medical images, public datasets along with segmentation masks are selected for training data synthesis.
As shown in Fig.~\ref{fig:workflow} (a), $\left \{ y^S, m^S \right \}$ are a sample of the public dataset $\left \{Y^S, M^S \right \}$, and $x^S \in X^S$ is a low-quality image synthesized from $y^S$ using degradation models.
Such that in the training data, $x^S$ is companied by an enhancement reference $y^S$ and structure guidance $m^S$.

Accordingly, we compose an enhancement network $\Phi ^S$ with structure constraints, where $\Phi ^S$ consists of an enhancement branch built by a U-Net~\cite{ronneberger2015u} architecture and an extra decoder for structure preservation. 
In detail, $x^S$ is embedded by an encoder $\mathcal{E}$, and the enhanced image is reconstructed by a decoder $\mathcal{D}$, whose layers are skip-connected with $\mathcal{E}$.
The extra decoder $\mathcal{S}$ is layer-level attached to $\mathcal{D}$ to impost structure preservation by predicting the segmentation mask $m^S$.

The enhancement loss of $\Phi ^S$ is formulated as
\begin{equation}
\mathcal{L}^S_1(\mathcal{E},\mathcal{D})=\mathbb{E} \left \|\hat{y}^S-y^S \right \|_1,
\label{eq:ls1}
\end{equation}
where $\hat{y}^S=\mathcal{D}(\mathcal{E}(x^S))$ denotes the enhanced sample.

And the structure-preserving loss of $\Phi ^S$ is calculated by
\begin{equation}
\mathcal{L}^S_{CE}(\mathcal{E},\mathcal{D},\mathcal{S})=\mathbb{E}\left [ - {\textstyle \sum_{c=1}^{C}m^S_c}\log{(\hat{m}^S_c)}\right ],
\label{eq:lsce}
\end{equation}
where $\hat{m}^S=\mathcal{S}(\mathcal{D}(\mathcal{E}(x^S)))$ represents the predicted segmentation masks.
$C$ denotes the number of segmentation categories, and $c$ represents the category.

The total optimization objective of $\Phi ^S$ is given by
\begin{equation}
\mathcal{L}^S(\mathcal{E},\mathcal{D},\mathcal{S})=\mathcal{L}^S_1(\mathcal{E},\mathcal{D})+\lambda _S \mathcal{L}^S_{CE}(\mathcal{E},\mathcal{D},\mathcal{S}),
\label{eq:overall}
\end{equation}
where $\lambda _S$ is the weight to balance the enhancement loss and structure-preserving loss, and is set to 0.3 here.

Leveraging public datasets and degradation models, training data are synthesized for enhancement networks. 
To ensure structure preservation in the enhancement, segmentation prediction is joined with the enhancement.
Once the source model $\Phi ^S$ is well-trained based on the source domain of training data, the source data is no longer available in the subsequent steps.

\subsection{Source-free unsupervised domain adaptive medical image enhancement}
SFUDA fine-tunes pre-trained models with unannotated target data exclusively, which not only enables further model optimization using clinical data, but also eliminates concerns related to privacy and cost associated with data collection and centralization.
Despite this potential, the application of SFUDA to medical image enhancement remains unexplored. 
{SAME bridges this gap by adopting knowledge distillation, which involves a teacher-student model to adapt the source model to the target domain, as exhibited in Fig.~\ref{fig:workflow} (b).
}

{The pseudocode of the distillation process is showcased in Algorithm 1.}
Both the teacher model $\Phi ^T_\alpha$ and the student model $\Phi ^T_\beta$ are initialized by the pre-trained source model $\Phi ^S$.
The teacher model $\Phi ^T_\alpha$ is fed with low-quality samples $X^T$ from the target domain to acquire enhancement results $Y^T_\alpha$ and segmentation results $M^T_\alpha$. 
A picker (Sec.~\ref{sec:picker}) is then introduced to select the desired enhanced and segmented outcomes as pseudo-labels to construct a proxy dataset $\{X^T_\Omega, Y^T_\Omega,M^T_\Omega \}$, which are used to further optimize the enhancement model.

{Then to facilitate the distillation, random variances of brightness, contrast, and color have been slightly applied to perturb the input space for the student model $\Phi ^T_\beta$.
The selected test sample $x^T_\Omega \in X^T_\Omega$ is subjected to variations to acquire the perturbed sample $\tilde{x}^T$.}
The student model $\Phi ^T_\beta$ loads $\tilde{x}^T$ as the input and the corresponding $\{y^T_\Omega,m^T_\Omega\} \in \{ Y^T_\Omega,M^T_\Omega \}$ as the enhancement and segmentation supervision to calculate optimization gradient.
The optimization objective of $\Phi ^T_\beta$ is defined as 
\begin{equation}
\mathcal{L}^T_\beta(\mathcal{E_\beta},\mathcal{D_\beta},\mathcal{S_\beta})=\mathcal{L}^T_1+\lambda_{CE} \mathcal{L}^T_{CE},
\label{eq:overall}
\end{equation}
where $\lambda_{CE}$ is set to 0.3.

The enhancement loss $\mathcal{L}^T_1$ is given by
\begin{equation}
\mathcal{L}^T_1(\mathcal{E_\beta},\mathcal{D_\beta})=\mathbb{E} \left \|y^T_\beta-y^T_\alpha \right \|_1,
\label{eq:lt1}
\end{equation}
where $y^T_\alpha$ and $y^T_\beta$ are the images enhanced by $\Phi ^T_\alpha$ and $\Phi ^T_\beta$.

While the structure-preserving loss $\mathcal{L}^T_{CE}$ is formulated as
\begin{equation}
\mathcal{L}^T_{CE}(\mathcal{E_\beta},\mathcal{D_\beta},\mathcal{S_\beta})=\mathbb{E}\left [ - {\textstyle \sum_{c=1}^{C}m^T_{\alpha c}}\log{(m^T_{\beta c})}\right ],
\label{eq:ltce}
\end{equation}
where $m^T_\alpha$ and $m^T_\beta$ are the masks predicted by $\Phi ^T_\alpha$ and $\Phi ^T_\beta$.

The parameters of $\Phi ^T_\alpha$ are updated by the exponential moving average (EMA) of student model weights in each training iteration, and
the pseudo-labels are also updated following $\Phi ^T_\alpha$.
{Repeat the above steps, until the optimization gradient converges. 
By doing so, the teacher-student model is adapted to the target domain, enabling the generation of desired enhanced images for the target sample $X^T$.}

\begin{algorithm}[tbp]
\caption{Knowledge distillation process for SAME on target domains}
\begin{algorithmic}[1]
    \Require 
    Source model $\Phi ^S$, Enhancement pseudo-label picker, Target data $X^T$.
    \Ensure
    Teacher model $\Phi ^T_\alpha$, Student model $\Phi ^T_\beta$
        \State Initialize a teacher model $\Phi ^T_\alpha$ and a student model $\Phi ^T_\beta$ with $\Phi ^S$.
        \While{not converged}
            \State  From target sample $x^T \in X^T$, $\Phi ^T_\alpha$ produces enhanced sample $y^T_\alpha \in Y^T_\alpha$ and segmentation mask $m^T_\alpha \in M^T_\alpha$.
            \State From $\{Y^T_\alpha,M^T_\alpha \}$, the picker selects enhancement pseudo-labels to construct proxy dataset $\{X^T_\Omega, Y^T_\Omega,M^T_\Omega \}$.
            \State Update $\Phi ^T_\beta$ based on $\{X^T_\Omega, Y^T_\Omega,M^T_\Omega \}$ using $\mathcal{L}^T_\beta$. Specifically, The input $\tilde{x}^T$ for $\Phi ^T_\beta$ is obtained by applying a random perturbation to $x^T_\Omega \in X^T_\Omega$. 
            \State Update $\Phi ^T_\alpha$ using the EMA of the weights from $\Phi ^T_\beta$.
        \EndWhile
\end{algorithmic}
\end{algorithm}

\subsection{Enhancement pseudo-label picker}
\label{sec:picker}
Efficient selection of appropriate pseudo-labels is essential for implementing the aforementioned knowledge distillation process.
Unlike explicit category labels in classification and segmentation tasks, evaluating the quality of enhancement inference is more challenging.
Enlighten by the medical image quality assessment and shape priors for segmentation~\cite{luo2020shape}, an image quality assessor (IQA) and an irregular structure detector (ISD) are constructed to cooperatively pick suitable pseudo-labels for $\Phi ^T_\beta$.

\begin{figure}[tbp]
\begin{centering}
\includegraphics[width=1\linewidth]{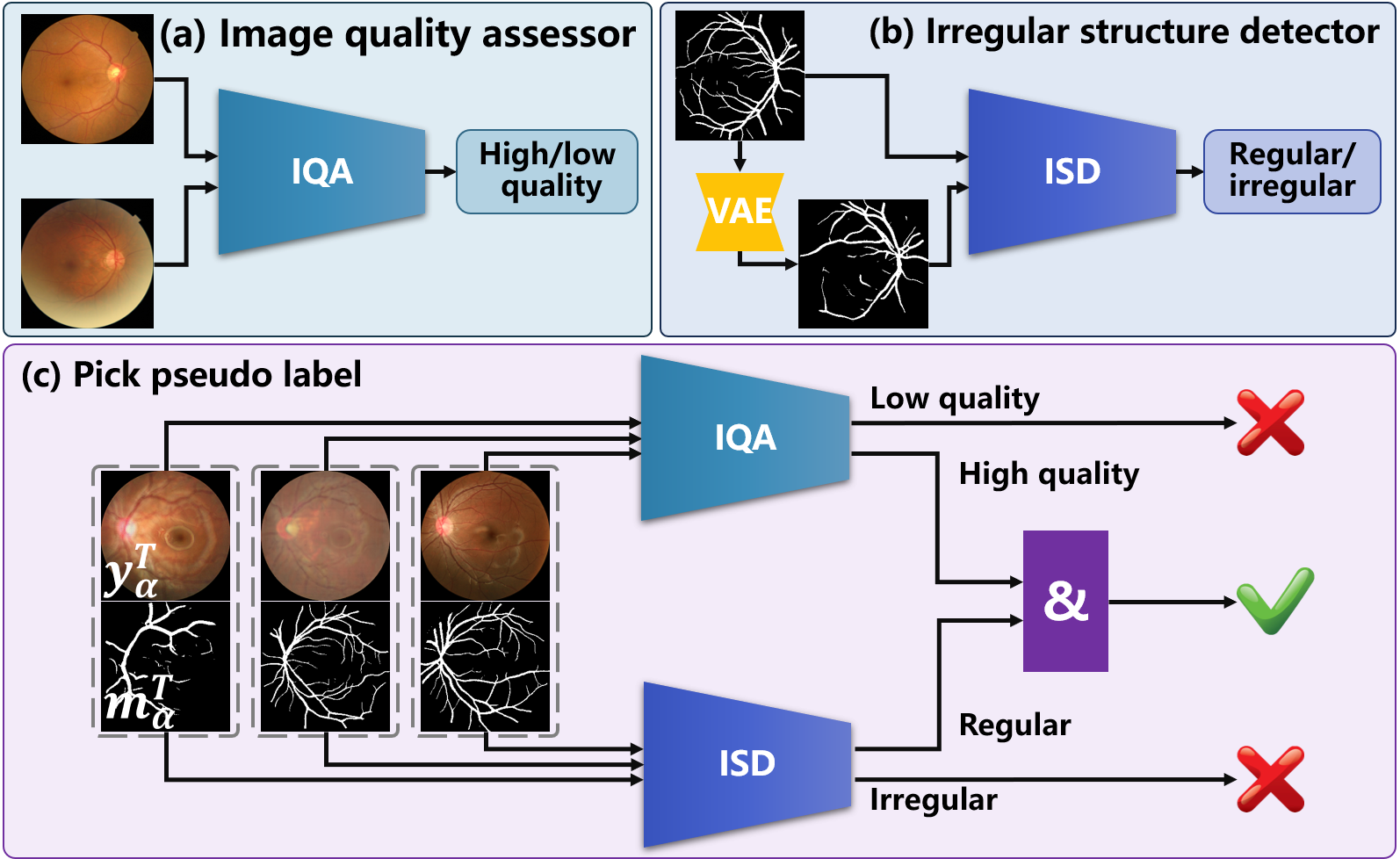}
\par\end{centering}
\caption{Pseudo-label picker construction. 
(a) An image quality assessor is developed using high- and low-quality samples. 
(b) An irregular structure detector is built by discriminating the ground truth masks of segmentation and masks reconstructed by a VAE.
(c) The quality assessor and irregularity detector cooperate to pick enhancement pseudo-labels for SFUDA, where a sample enhanced by the target teacher model will only be picked as a pseudo-label if it is determined to have high-quality and regular structures.}
\label{fig:label}
\end{figure}

As demonstrated in Fig.~\ref{fig:label} (a), the IQA is implemented using a classifier trained to identify the high- and low-quality medical image samples.
The assessment network reported in~\cite{fu2019evaluation} can be employed as the IQA, where the optimization objective is defined as
\begin{equation}
\mathcal{L}_{IQA}=\mathbb{E}\left [\varphi (IQA(x_q)-q) \right ].
\label{eq:IQA}
\end{equation}
where $\{x_q,q\}$ is the images sample and quality label. 
$\varphi (\cdot )$ refers to cross-entropy loss.

On the other hand, the ISD is developed in the adversarial training of a variational autoencoder generative adversarial network (VAE-GAN). 
The adversarial loss is given by 
\begin{equation}
\mathcal{L}_{VAE-GAN}=D(m)+\mathrm{log} (1-D(\hat{m})).
\label{eq:ISD}
\end{equation}
where $m$ is a ground truth mask of segmentation, $\hat{m}$ is a mask generated by the VAE-GAN based on vectors randomly sampled from latent space.
As shown in Fig.~\ref{fig:label} (b), a minimax game is played between the VAE and a discriminator $D$, whose responsibility is to distinguish between $m$ and $\hat{m}$ in the adversarial training.
Thus $D$ can be used as the ISD to detect irregular structures.

{
In order to ensure the picker's generalizability across different domains and facilitate the SFUDA process, public datasets have been collected to train both the IAQ and the ISD.
The IAQ is trained using a publicly available large dataset EyeQ}\footnote{https://github.com/HzFu/EyeQ/}{, which consists of 28,792 samples. 
The large volume and discrepancy of EyeQ enable the IQA to robustly perform in unfamiliar domains.
Similarly, the ISD has been trained utilizing public segmentation datasets, namely DRIVE}\footnote{http://www.isi.uu.nl/Research/Databases/DRIVE/}{, AVRDB}\footnote{https://data.mendeley.com/datasets/3csr652p9y/}{, and DR HAGIS}\footnote{https://pubmed.ncbi.nlm.nih.gov/28217714/}{. The ISD is optimized by identifying the generated structure masks from the ground truth ones of these datasets. Since the structures present in medical images tend to remain consistent across domains, the ISD is considered to be domain-agnostic.
As a result, the picker can be directly applied to target domains without the need for further tuning using target-specific data.
}

{During the adaptation phase, the IQA and the ISD are frozen and collaborate to pick appropriate pseudo-labels for $\Phi ^T_\beta$ from the outcomes of $\Phi ^T_\alpha$.}
A pseudo-label is picked only if $y^T_\alpha$ is determined to be of high-quality and $m^T_\alpha$ is identified as a regular structure. 
Accordingly, appropriate pseudo-labels are selected for achieving SFUDA.

\section{Experiments}
To demonstrate the performance of the proposed SAME, extensive experiments were carried out. 
Comparisons with SOTA medical image enhancement algorithms were presented on data dependency, enhancement performance, and computational complexity.
Segmentation and diagnosis tasks were also conducted to validate the benefits of the enhancement.
Setting analysis of source model training and ablation studies were then implemented to verify the effectiveness.

\subsection{Experimental Settings}
Ten SOTA enhancement algorithms and ten datasets from three medical image modalities were collected to verify the performance of the proposed SAME. 
The evaluation was conducted on three tasks to comprehensively interpret the advantages of SAME.

\subsubsection{Baselines}
Comparisons with the SOTA medical image enhancement algorithms are presented to demonstrate the advantages of SAME.

For fundus photography, eight algorithms designed to enhance fundus images were introduced as baselines. 
The model of RFormer~\cite{RFormer} was trained on paired clinical samples using Transformer and is publicly available from the authors' homepage.
Based on unpaired data, StillGAN~\cite{ma2021structure} and I-SECRET~\cite{cheng2021secret} were developed to improve fundus image quality.
ArcNet~\cite{li2022annotation} and MAGE-Net~\cite{guo2023bridging} access test data during training to adapt the enhancement model to target data.
CofeNet~\cite{shen2020modeling} utilizes segmentation to boost structure preservation in the enhancement.
PCE-Net~\cite{liu2022degradation} and SCR-Net~\cite{li2022structure}  attempt to learn robust enhancement models respective to imaging interference and cataracts by constraining feature consistency.

For OCT and ultrasound images,  three algorithms for medical or OCT image enhancement were implemented. As HDcycleGAN~\cite{manakov2019noise} and StillGAN~\cite{ma2021structure} were developed for any medical images, they were selected to enhance OCT and ultrasound images. The diffusion-based denoising algorithm for OCT, ODDM~\cite{hu2022unsupervised} has also been resorted as the baseline.

\begin{table}[htbp]
\footnotesize
\centering
\caption{Experiment tasks, datasets, and evaluation metrics}
\label{tab:datasets} 
\renewcommand{\arraystretch}{1}
\begin{tabular}{m{1.5cm} | m{1.9cm} | m{2cm} | m{1.7cm}}
\hline
\hspace{0.8em}\textbf{Modality} & \hspace{0.2em}\textbf{Training data} & \hspace{1.1em}\textbf{Test data} & \textbf{Tasks(Metrics)}\\
\hline
Fundus photography & DRIVE, AVR-DB, DR HAGIS & RF, FIQ, RCF & \multirow{3}{*}{\parbox{1.7cm}{Enhancement (SSIM, PSNR) / Segmentation (DICE, IOU)}}\\
\cline{1-3}
OCT & Clear samples in EHFU & Paired data in EHFU, A2A & \\
 \cline{1-3}
Ultrasound & Clear samples in SUStecH & Paired data in SUStecH & \\
\hline
Fundus photography & Clear samples in Fundus-iSee & Low-quality ones in Fundus-iSee &
Diagnosis (F1-score, Ckappa)\\
\hline
\end{tabular}%
\end{table}

\begin{table*}[tbp]
\footnotesize
\centering
\caption{Comparisons with SOTA fundus image enhancement algorithms on data dependency, enhancement performance, and computational complexity}
\label{tab:comparison_SOTA} 
\renewcommand{\arraystretch}{0.9}
\begin{threeparttable}
\begin{tabular}{p{1.9cm} | p{0.3cm}<{\centering} p{0.3cm}<{\centering} p{0.3cm}<{\centering}| p{0.7cm}<{\centering} p{0.7cm}<{\centering} p{0.7cm}<{\centering} p{0.7cm}<{\centering}| p{0.7cm}<{\centering} p{0.7cm}<{\centering} p{0.7cm}<{\centering} p{0.7cm}<{\centering}| p{0.8cm}<{\centering} p{0.8cm}<{\centering} p{0.9cm}<{\centering}}
\hline
\multirow{2}{*}{Algorithms} & \multicolumn{3}{c|}{Dependency*} & \multicolumn{4}{c|}{SSIM**} &  \multicolumn{4}{c|}{PSNR**} & Costs & Training & Inference\\
\cline{2-15}
& \textsl{PD} & \textsl{UD} & \textsl{VT} & RF & FIQ & RCF & Avg. & RF & FIQ & RCF & Avg. & (GMac) & (Hours) & (Seconds)\\
\hline
\multirow{2}{*}{RFormer~\cite{RFormer}} 
&  \multirow{2}{*}{\textcolor{red}{$\star$}} & &
& \textbf{0.873}& 0.788& 0.728& 0.796& \textbf{28.32}& 16.61& 17.14& 20.69
& 45.46& --& 0.16\\
&&&& \raisebox{0.2em}{\scriptsize (0.050)} &  \raisebox{0.2em}{\scriptsize (0.070)} & \raisebox{0.2em}{\scriptsize (0.064)} & \raisebox{0.2em}{\scriptsize (0.059)} 
&  \raisebox{0.2em}{\scriptsize (2.70)}  & \raisebox{0.2em}{\scriptsize (2.44)} &  \raisebox{0.2em}{\scriptsize (2.26)} & \raisebox{0.2em}{\scriptsize (5.40)} 
&&&\\

\multirow{2}{*}{StillGAN~\cite{ma2021structure}} 
& & \multirow{2}{*}{\textcolor{red}{$\star$}} &
& 0.760& 0.871& 0.748& 0.793& \textbf{24.17}& 21.44& 18.24& 21.28
& 67.12& 51.71& 0.20\\
&&&& \raisebox{0.2em}{\scriptsize (0.082)} &  \raisebox{0.2em}{\scriptsize (0.071)} & \raisebox{0.2em}{\scriptsize (0.054)} & \raisebox{0.2em}{\scriptsize (0.055)} 
&  \raisebox{0.2em}{\scriptsize (3.72)}  & \raisebox{0.2em}{\scriptsize (5.98)} &  \raisebox{0.2em}{\scriptsize (6.27)} & \raisebox{0.2em}{\scriptsize (2.42)} 
&&&\\

\multirow{2}{*}{I-SECRET~\cite{cheng2021secret}} 
& & \multirow{2}{*}{\textcolor{red}{$\star$}} & \multirow{2}{*}{\textcolor{red}{$\star$}} 
& 0.756& 0.868& 0.750& 0.791& 22.90& 21.32& 18.49& 20.90
& 56.88& 16.88& 0.18\\
&&&& \raisebox{0.2em}{\scriptsize (0.062)} &  \raisebox{0.2em}{\scriptsize (0.051)} & \raisebox{0.2em}{\scriptsize (0.074)} & \raisebox{0.2em}{\scriptsize (0.054)} 
&  \raisebox{0.2em}{\scriptsize (5.55)}  & \raisebox{0.2em}{\scriptsize (7.43)} &  \raisebox{0.2em}{\scriptsize (5.08)} & \raisebox{0.2em}{\scriptsize (1.82)} 
&&&\\

\multirow{2}{*}{ArcNet~\cite{li2022annotation}} 
& & & \multirow{2}{*}{\textcolor{red}{$\star$}} 
& 0.758& 0.868& 0.760& 0.795& 23.11& 21.51& 18.36& 20.99
& \textbf{18.16}& 6.84& \textbf{0.10}\\
&&&& \raisebox{0.2em}{\scriptsize (0.053)} &  \raisebox{0.2em}{\scriptsize (0.053)} & \raisebox{0.2em}{\scriptsize (0.064)} & \raisebox{0.2em}{\scriptsize (0.051)} 
&  \raisebox{0.2em}{\scriptsize (6.99)}  & \raisebox{0.2em}{\scriptsize (12.38)} &  \raisebox{0.2em}{\scriptsize (3.43)} & \raisebox{0.2em}{\scriptsize (1.97)} 
&&&\\

\multirow{2}{*}{MAGE-Net~\cite{guo2023bridging}} 
& & & \multirow{2}{*}{\textcolor{red}{$\star$}} 
& 0.760& 0.861& 0.762& 0.794& 23.14& 21.64& 18.12& 20.96
& 854.32& 16.23& 0.27\\
&&&& \raisebox{0.2em}{\scriptsize (0.046)} &  \raisebox{0.2em}{\scriptsize (0.035)} & \raisebox{0.2em}{\scriptsize (0.061)} & \raisebox{0.2em}{\scriptsize (0.471)} 
&  \raisebox{0.2em}{\scriptsize (2.19)}  & \raisebox{0.2em}{\scriptsize (2.09)} &  \raisebox{0.2em}{\scriptsize (2.17)} & \raisebox{0.2em}{\scriptsize (2.10)} 
&&&\\

\multirow{2}{*}{CofeNet~\cite{shen2020modeling}} 
& & & 
& 0.717& 0.838& 0.744& 0.766& 19.10& 20.64& 17.83& 19.19
& 67.50& 19.76& 0.19\\
&&&& \raisebox{0.2em}{\scriptsize (0.042)} &  \raisebox{0.2em}{\scriptsize (0.041)} & \raisebox{0.2em}{\scriptsize (0.043)} & \raisebox{0.2em}{\scriptsize (0.052)} 
&  \raisebox{0.2em}{\scriptsize (3.25)}  & \raisebox{0.2em}{\scriptsize (5.14)} &  \raisebox{0.2em}{\scriptsize (4.45)} & \raisebox{0.2em}{\scriptsize (1.15)} 
&&&\\

\multirow{2}{*}{PCE-Net~\cite{liu2022degradation}} 
&  & &  
& 0.745& 0.872& 0.736& 0.784& 18.86& \textbf{23.09}& 17.30& 19.75
& 85.29& 12.90 &0.25 \\
&&&& \raisebox{0.2em}{\scriptsize (0.032)} &  \raisebox{0.2em}{\scriptsize (0.031)} & \raisebox{0.2em}{\scriptsize (0.035)} & \raisebox{0.2em}{\scriptsize (0.062)} 
&  \raisebox{0.2em}{\scriptsize (3.32)}  & \raisebox{0.2em}{\scriptsize (7.03)} &  \raisebox{0.2em}{\scriptsize (5.64)} & \raisebox{0.2em}{\scriptsize (2.45)} 
&&&\\

\multirow{2}{*}{SCR-Net~\cite{li2022structure}} 
&  & &  
& 0.752& 0.871 & \textbf{0.773}& 0.799& 18.08& 21.56& 18.39& 19.34
& 34.80& 3.43& 0.14\\
&&&& \raisebox{0.2em}{\scriptsize (0.033)} &  \raisebox{0.2em}{\scriptsize (0.051)} & \raisebox{0.2em}{\scriptsize (0.069)} & \raisebox{0.2em}{\scriptsize (0.052)} 
&  \raisebox{0.2em}{\scriptsize (3.77)}  & \raisebox{0.2em}{\scriptsize (6.44)} &  \raisebox{0.2em}{\scriptsize (5.99)} & \raisebox{0.2em}{\scriptsize (1.57)} 
&&&\\

\hline
\multirow{2}{*}{SAME-source} 
&  & &  
&0.750 & 0.862 & 0.761& 0.791
&22.18 &21.67 &18.19 & 20.68
& 34.02 & 3.19 & 0.13 \\
&&&& \raisebox{0.2em}{\scriptsize (0.047)} &  \raisebox{0.2em}{\scriptsize (0.028)} & \raisebox{0.2em}{\scriptsize (0.061)} & \raisebox{0.2em}{\scriptsize (0.050)} 
&  \raisebox{0.2em}{\scriptsize (3.78)}  & \raisebox{0.2em}{\scriptsize (4.46)} &  \raisebox{0.2em}{\scriptsize (5.11)} & \raisebox{0.2em}{\scriptsize (1.77)} 
&&&\\

\multirow{2}{*}{SAME (ours)} 
& & &  
&\textbf{0.770} & \textbf{0.873} &0.771& \textbf{0.805}
&23.95 &23.06 &\textbf{18.88} & \textbf{21.96}
& 68.04 & \textbf{2.67}*** & 0.13 \\
&&&& \raisebox{0.2em}{\scriptsize (0.044)} &  \raisebox{0.2em}{\scriptsize (0.023)} & \raisebox{0.2em}{\scriptsize (0.051)} & \raisebox{0.2em}{\scriptsize (0.048)} 
&  \raisebox{0.2em}{\scriptsize (3.55)}  & \raisebox{0.2em}{\scriptsize (4.22)} &  \raisebox{0.2em}{\scriptsize (4.98)} & \raisebox{0.2em}{\scriptsize (2.21)} 
&&&\\

\hline
\end{tabular}%
\begin{tablenotes}
 \scriptsize
 \item{*} Dependency on high-low quality paired clinical data (\textsl{PD}), unpaired clinical data (\textsl{UD}), and visiting test data during training (\textsl{VT}) are indicated by \textcolor{red}{$\star$}. 
 ** Besides RFormer, the second top result on RF is also highlighted.
{*** This training time refers to the time taken for the SFUDA process.}
\end{tablenotes}
\end{threeparttable}
\end{table*}

\subsubsection{Datasets}
As summarized in Table~\ref{tab:datasets}, three medical image modalities were leveraged in the experiments:

\noindent $\triangleright$ \textbf{Fundus photography}

\noindent \textbf{DRIVE}: 40 clear fundus images annotated by vessel masks.

\noindent \textbf{AVRDB}: 100 clear fundus images for vessel segmentation.

\noindent \textbf{DR HAGIS}: 39 clear fundus images for vessel segmentation.

\noindent \textbf{RF}\footnote{https://github.com/dengzhuo-AI/Real-Fundus/releases/download/v.1.0.0/ Real\_Fundus.zip}: 120 high-low quality paired fundus image samples.

\noindent \textbf{FIQ}: a fundus dataset containing 196 low-high quality image pairs collected from Shenzhen Kangning Hospital.

\noindent \textbf{RCF}: a fundus dataset collected from Peking University Third Hospital, consisting of 26 fundus images after cataract surgery corresponding to the ones before surgery.

\noindent \textbf{Fundus-iSee}: a fundus dataset including 10,000 images (2,669 low-quality samples primarily affected by cataracts and 7,331 high-quality ones), sorted into five categories according to fundus status.

\noindent $\triangleright$ \textbf{OCT}

\noindent \textbf{EHFU OCT}: an OCT dataset for retinal layer segmentation composed of 157 high-quality samples and 86 high-low quality pairs collected from the Eye and ENT Hospital of Fudan University.

\noindent \textbf{A2A SD-OCT}\footnote{https://people.duke.edu/~sf59/Fang\_BOE\_2012.htm}: 17 high-low quality SD-OCT image pairs.

\noindent $\triangleright$ \textbf{Ultrasound}

\noindent \textbf{SUStecH}: an ultrasound dataset containing 5740 images for articular cartilage segmentation provided by Southern University of Science and Technology Hospital, from which 1000 high-quality samples and 862 high-low quality pairs were used in the experiment.

\subsubsection{Implementation and evaluation metrics}
The tasks of enhancement, segmentation, and diagnosis were performed to understand the benefits of SAME.

{For synthetic data-based algorithms, paired training data were generated by applying degradations modeled in \cite{shen2020modeling,li2022annotation}. Multiple degraded samples were synthesized from each high-quality sample by randomly varying the degradation parameters. For instance, 16 degraded samples were synthesized from each high-quality sample in DRIVE.
On the other hand, unpaired data-based algorithms were trained using clear samples from the public training data and low-quality test data. To ensure a fair experiment, data augmentation techniques such as flipping, cropping, and rotation were employed on the unpaired training data to match the training data size of synthetic data-based algorithms.
}
Test data were also accessed by I-SECRET~\cite{cheng2021secret}, ArcNet~\cite{li2022annotation}, and MAGE-Net~\cite{guo2023bridging} to execute DA.
SAME is initialized by the source model from synthetic training data, and optimized with the test data using SFUDA. 
Public code was utilized to implement comparative algorithms. 
SAME is reformed from a U-Net architecture.

The input image size for training was 256 $\times$ 256 and the batch size was 8. 
The training data were loaded with a random scale among \{286, 306, 326, 346\}, and then cropped to the size of 256.
The model was trained by the Adam optimizer for 150 epochs with an initial learning rate of 0.001 and 50 epochs with the learning rate gradually decaying to 0. 

The enhancement performance was quantified by structural similarity (SSIM) and the peak signal-to-noise ratio (PSNR).
U-Net and ResNet-50 were respectively employed to construct a segmentation and a diagnosis model based on DRIVE and Fundus-iSee to fulfill the downstream tasks.
The segmentation improvement achieved by the enhancement was quantified by calculating the intersection over union (IoU) and the Dice coefficient between the segmentation results of the low-quality images and their corresponding enhanced versions.
On the other hand, F1-score and Cohen's kappa (Ckappa) were computed to compare the diagnosis performance on the low-quality, clear, and enhanced images.

\subsection{Comparisons with SOTA algorithms}
{A comparison is conducted against SOTA algorithms in medical image enhancement and SFUDA paradigms, to assess the advantages of SAME in fundus image enhancement, as well as its impact on downstream tasks such as segmentation and diagnosis.}

\subsubsection{Comparisons with fundus image enhancement algorithms}
\begin{figure*}[htbp]
\begin{centering}
\includegraphics[width=0.9\linewidth]{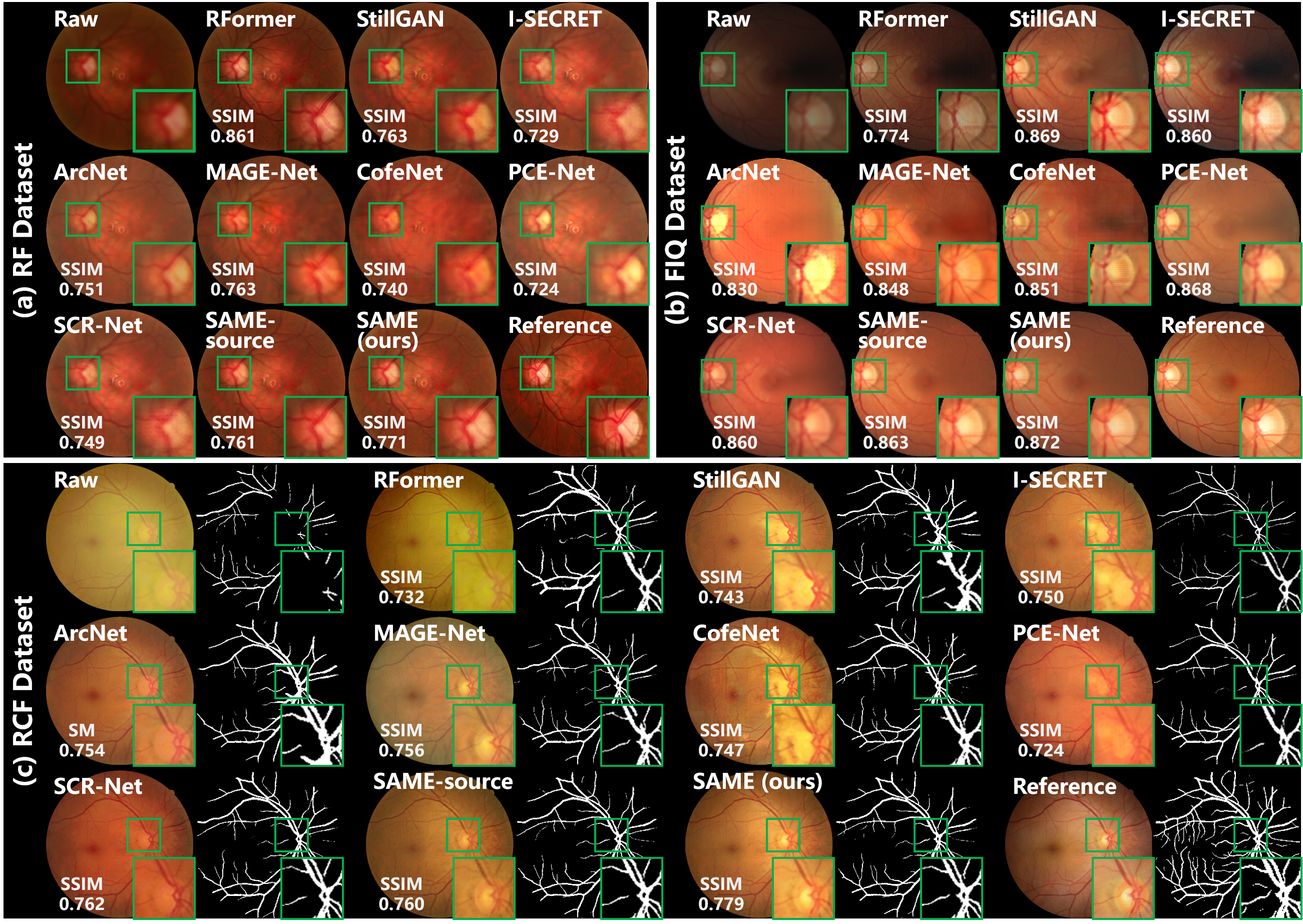}
\par\end{centering}
\caption{Visualized comparison of enhancement and segmentation with SOTA algorithms. 
{(a) and (d) showcase the enhancement results of RF and FIQ, while both enhancement and segmentation results of RCF are presented in (c).
SAME improves the source model in contrast, structural clarity, and color authenticity of the enhanced images, and outperforms the comparative algorithms in the consistency with the reference images. 
}
}
\label{fig:comparison}
\end{figure*}

Table~\ref{tab:comparison_SOTA} summarizes the comparison with SOAT fundus image enhancement algorithms in qualitative data dependency, quantitative enhancement, and computational complexity. 
The mean and standard deviation of the enhancement metrics for SSIM and PSNR are presented, and the performance results of both the source initialized SAME (SAME-source) and the source-free unsupervised domain adapted SAME (ours) are provided. 
Figure~\ref{fig:comparison} provides visualized comparisons with the algorithms. 
Notably, only DRIVE is used to synthesize the paired training data in this comparison.

\textbf{Data dependency. }
Developing an efficient enhancement model depends on redundant training data. 
As shown in Table~\ref{tab:comparison_SOTA}, different solutions have been designed to achieve the training requirement.

RFormer~\cite{RFormer} is trained with the RF dataset, which was built in clinics by collecting plenty of high-low quality image pairs.
Unfortunately, collecting paired data is extremely costly and troublesome, and causes privacy concerns.
To alleviate the challenges in data collection, unpaired training algorithms have been developed.
StillGAN~\cite{ma2021structure} and I-SECRET~\cite{cheng2021secret} are enabled to learn enhancement models based on unpaired training data.
However, unpaired training still requires sufficient clinical data and has privacy risks.
Synthetic training data were thus leveraged to learn enhancement models. 
Considering the domain shifts between synthetic and real-world data, ArcNet~\cite{li2022annotation} and MAGE-Net~\cite{guo2023bridging} access the to-be-enhanced test data during the training phase to conduct domain adaptation.
On the other hand, CofeNet~\cite{shen2020modeling} directly applied the model learned from synthetic to enhance clinical images.
To compress the privacy risks of test data access during training, PCE-Net~\cite{liu2022degradation} and SCR-Net~\cite{li2022structure}  impose model generalizability by constraining feature consistency.
While PCE-Net~\cite{liu2022degradation} and SCR-Net~\cite{li2022structure}  circumvent the privacy issue and achieve remarkable performance, they are incapable of optimizing the models further with test data.

The training on source data promises SAME a reasonable initialization.
And SFUDA endows SAME to further optimize the enhancement model only based on test data.
Therefore, SAME not only makes full use of the data to further optimize the model, but also avoids the privacy issue of accessing training and testing data simultaneously.

\textbf{Enhancement performance. }
Table~\ref{tab:comparison_SOTA} summarizes the enhancement performance on the datasets of RF, FIQ, and RCF, and the average value of the three datasets.
Fig.~\ref{fig:comparison} shows the enhanced images by various algorithms, as well as the raw and reference samples in the three datasets.

As RFormer~\cite{RFormer} was designed for and trained with specifically RF dataset, it is no surprise that RFormer~\cite{RFormer} behaves dominantly on RF data. 
But when it comes to FIQ and RCF, the mediocre performance of RFormer~\cite{RFormer} is observed from Table~\ref{tab:comparison_SOTA}.
Using unpaired training data, StillGAN~\cite{ma2021structure} and I-SECRET~\cite{cheng2021secret} provide decent enhancement performance in the metrics of SSIM and PSNR. 
But the images enhanced by StillGAN~\cite{ma2021structure} and I-SECRET~\cite{cheng2021secret} suffer from uneven color according to Fig.~\ref{fig:comparison}.
Test data were visited by ArcNet~\cite{li2022annotation} and MAGE-Net~\cite{guo2023bridging} to carry out domain adaptation, which generalizes the model from synthetic to real-world data.
Due to the GAN-based framework, the training of ArcNet~\cite{li2022annotation} is a delicate procedure and may result in fluctuating performance as shown in Fig.~\ref{fig:comparison}.
CofeNet~\cite{shen2020modeling} ignores the domain shifts between synthetic and real-world data, such that inferior results are obtained.
Through the constraint of feature consistency, PCE-Net~\cite{liu2022degradation} and SCR-Net~\cite{li2022structure}  are equipped with generalizability, which allows them to outperform CofeNet~\cite{shen2020modeling} under the same data dependency.
Additionally, PCE-Net~\cite{liu2022degradation} and SCR-Net~\cite{li2022structure} exhibit exceptional performance on FIQ and RCF, respectively, as they are tailored to address imaging interference and cataracts in fundus photography.
{However, Fig.~\ref{fig:comparison} reveals that images enhanced by PCE-Net and SCR-Net inherit specific color styles that are independent of the raw images, leading to a less favorable evaluation in terms of PSNR in Table~\ref{tab:comparison_SOTA}.}

In the source domain of synthetic training data, structure consistency is leveraged to facilitate SAME to learn a robust source model, which achieves decent initialization on all datasets.
Additionally, SAME further optimizes the enhancement model with only the test data, leading to an exceptional average performance on various datasets.
{
As shown in Fig.~\ref{fig:comparison}, SAME improves the source model in image contrast, structural clarity, and color fidelity of enhanced images, which are more consistent with reference images.
}

\begin{table*}[!t]
\footnotesize
\centering
\caption{Comparisons on down-stream tasks of fundus analysis}
\label{tab:comparison_seg} 
\renewcommand{\arraystretch}{1}
\begin{tabular}{p{2.2cm} | p{0.8cm}<{\centering} p{0.8cm}<{\centering} p{0.8cm}<{\centering} p{0.8cm}<{\centering}| p{0.8cm}<{\centering} p{0.8cm}<{\centering} p{0.8cm}<{\centering} p{0.8cm}<{\centering}| p{1.0cm}<{\centering} p{1.0cm}<{\centering}}
\hline
\multirow{3}{*}{Algorithms} & \multicolumn{8}{c|}{Segmentation} & \multicolumn{2}{c}{Diagnosis}\\
\cline{2-11}
& \multicolumn{4}{c|}{DICE} & \multicolumn{4}{c|}{IoU} & F1-score & Ckappa\\
\cline{2-11}
& RF & FIQ & RCF & Avg. & RF & FIQ & RCF & Avg. & \multicolumn{2}{c}{Fundus-iSee}\\
\hline
\multirow{2}{*}{Low-quality} & 0.469& 0.304& 0.518&  0.430
& 0.306& 0.179& 0.350& 0.278 & 0.730& 0.310\\
& \raisebox{0.2em}{\scriptsize (0.031)} &  \raisebox{0.2em}{\scriptsize (0.021)} & \raisebox{0.2em}{\scriptsize (0.042)} & \raisebox{0.2em}{\scriptsize (0.092)} 
&  \raisebox{0.2em}{\scriptsize (0.025)}  & \raisebox{0.2em}{\scriptsize (0.019)} &  \raisebox{0.2em}{\scriptsize (0.036)} & \raisebox{0.2em}{\scriptsize (0.072)} 
&&\\

\hline
\multirow{2}{*}{RFormer~\cite{RFormer}} & 0.577 & 0.580& 0.344& 0.500 
& 0.406 & 0.410& 0.194& 0.336 & 0.732& 0.370\\
& \raisebox{0.2em}{\scriptsize (0.042)} &  \raisebox{0.2em}{\scriptsize (0.039)} & \raisebox{0.2em}{\scriptsize (0.028)} & \raisebox{0.2em}{\scriptsize (0.111)} 
&  \raisebox{0.2em}{\scriptsize (0.032)}  & \raisebox{0.2em}{\scriptsize (0.052)} &  \raisebox{0.2em}{\scriptsize (0.047)} & \raisebox{0.2em}{\scriptsize (0.101)} 
&&\\

\multirow{2}{*}{StillGAN~\cite{ma2021structure}} & 0.559& 0.620& 0.544& 0.574 
& 0.388& 0.450& 0.373& 0.403 & 0.739& 0.348\\
& \raisebox{0.2em}{\scriptsize (0.048)} &  \raisebox{0.2em}{\scriptsize (0.051)} & \raisebox{0.2em}{\scriptsize (0.049)} & \raisebox{0.2em}{\scriptsize (0.033)} 
&  \raisebox{0.2em}{\scriptsize (0.027)}  & \raisebox{0.2em}{\scriptsize (0.038)} &  \raisebox{0.2em}{\scriptsize (0.029)} & \raisebox{0.2em}{\scriptsize (0.033)} 
&&\\

\multirow{2}{*}{I-SECRET~\cite{cheng2021secret}} & 0.563& 0.623& 0.541& 0.576
& 0.392& 0.453& 0.371& 0.405& 0.734& 0.382\\
& \raisebox{0.2em}{\scriptsize (0.049)} &  \raisebox{0.2em}{\scriptsize (0.052)} & \raisebox{0.2em}{\scriptsize (0.045)} & \raisebox{0.2em}{\scriptsize (0.034)} 
&  \raisebox{0.2em}{\scriptsize (0.028)}  & \raisebox{0.2em}{\scriptsize (0.039)} &  \raisebox{0.2em}{\scriptsize (0.030)} & \raisebox{0.2em}{\scriptsize (0.035)} 
&&\\

\multirow{2}{*}{ArcNet~\cite{li2022annotation}} & 0.569& 0.627& 0.572& 0.589
& 0.397& 0.456& 0.401& 0.418& 0.761& 0.428\\
& \raisebox{0.2em}{\scriptsize (0.050)} &  \raisebox{0.2em}{\scriptsize (0.057)} & \raisebox{0.2em}{\scriptsize (0.051)} & \raisebox{0.2em}{\scriptsize (0.027)} 
&  \raisebox{0.2em}{\scriptsize (0.035)}  & \raisebox{0.2em}{\scriptsize (0.042)} &  \raisebox{0.2em}{\scriptsize (0.036)} & \raisebox{0.2em}{\scriptsize (0.027)} 
&&\\

\multirow{2}{*}{MAGE-Net~\cite{guo2023bridging}} &0.588 & \textbf{0.646}& 0.585& 0.606 
& 0.417 & \textbf{0.472}& 0.414& 0.434& 0.753& 0.390\\
& \raisebox{0.2em}{\scriptsize (0.055)} &  \raisebox{0.2em}{\scriptsize (0.059)} & \raisebox{0.2em}{\scriptsize (0.061)} & \raisebox{0.2em}{\scriptsize (0.028)} 
&  \raisebox{0.2em}{\scriptsize (0.038)}  & \raisebox{0.2em}{\scriptsize (0.062)} &  \raisebox{0.2em}{\scriptsize (0.039)} & \raisebox{0.2em}{\scriptsize (0.027)} 
&&\\

\multirow{2}{*}{CofeNet~\cite{shen2020modeling}} & 0.551& 0.607& 0.529& 0.562 
& 0.381& 0.436& 0.360& 0.392& 0.754& 0.416\\
& \raisebox{0.2em}{\scriptsize (0.041)} &  \raisebox{0.2em}{\scriptsize (0.043)} & \raisebox{0.2em}{\scriptsize (0.039)} & \raisebox{0.2em}{\scriptsize (0.033)} 
&  \raisebox{0.2em}{\scriptsize (0.027)}  & \raisebox{0.2em}{\scriptsize (0.028)} &  \raisebox{0.2em}{\scriptsize (0.029)} & \raisebox{0.2em}{\scriptsize (0.032)} 
&&\\

\multirow{2}{*}{PCE-Net~\cite{liu2022degradation}} & 0.594& 0.634& 0.569 & 0.598 
& 0.422& 0.464& 0.397 & 0.427& 0.752& 0.410\\
& \raisebox{0.2em}{\scriptsize (0.054)} &  \raisebox{0.2em}{\scriptsize (0.064)} & \raisebox{0.2em}{\scriptsize (0.059)} & \raisebox{0.2em}{\scriptsize (0.027)} 
&  \raisebox{0.2em}{\scriptsize (0.039)}  & \raisebox{0.2em}{\scriptsize (0.045)} &  \raisebox{0.2em}{\scriptsize (0.038)} & \raisebox{0.2em}{\scriptsize (0.028)} 
&&\\

\multirow{2}{*}{SCR-Net~\cite{li2022structure}}  & 0.595& 0.613& \textbf{0.589}& 0.599 
& 0.424& 0.442& \textbf{0.417}& 0.428& \textbf{0.770}& 0.445\\
& \raisebox{0.2em}{\scriptsize (0.055)} &  \raisebox{0.2em}{\scriptsize (0.063)} & \raisebox{0.2em}{\scriptsize (0.057)} & \raisebox{0.2em}{\scriptsize (0.010)} 
&  \raisebox{0.2em}{\scriptsize (0.045)}  & \raisebox{0.2em}{\scriptsize (0.048)} &  \raisebox{0.2em}{\scriptsize (0.043)} & \raisebox{0.2em}{\scriptsize (0.011)} 
&&\\

\hline
\multirow{2}{*}{SAME-source} & 0.592 & 0.619 & 0.573 & 0.595
& 0.421 &0.448 & 0.402 & 0.424& 0.759 & 0.423\\
& \raisebox{0.2em}{\scriptsize (0.049)} &  \raisebox{0.2em}{\scriptsize (0.051)} & \raisebox{0.2em}{\scriptsize (0.055)} & \raisebox{0.2em}{\scriptsize (0.019)} 
&  \raisebox{0.2em}{\scriptsize (0.037)}  & \raisebox{0.2em}{\scriptsize (0.039)} &  \raisebox{0.2em}{\scriptsize (0.037)} & \raisebox{0.2em}{\scriptsize (0.019)} 
&&\\

\multirow{2}{*}{SAME (ours)} & \textbf{0.610} & 0.638 & 0.578 &\textbf{0.609 } 
& \textbf{0.451} &0.468 & 0.406 & \textbf{0.441}& 0.769 & \textbf{0.456}\\
& \raisebox{0.2em}{\scriptsize (0.048)} &  \raisebox{0.2em}{\scriptsize (0.053)} & \raisebox{0.2em}{\scriptsize (0.044)} & \raisebox{0.2em}{\scriptsize (0.025)} 
&  \raisebox{0.2em}{\scriptsize (0.036)}  & \raisebox{0.2em}{\scriptsize (0.048)} &  \raisebox{0.2em}{\scriptsize (0.041)} & \raisebox{0.2em}{\scriptsize (0.026)} 
&&\\

\hline
\end{tabular}%
\end{table*}


\textbf{Computational complexity analysis.}
Computational costs and time consumption are respectively quantified by multiply-accumulate operation (GMac) as well as training (Hours) and inference time (Seconds) to analyze the computational complexity of enhancement algorithms.

Please take note that the training time of RFormer~\cite{RFormer} is not provided, since it was performed with the publicly available pre-trained model.
In the case of SAME, the enhancement model is initialized with the source model learned from the training data and then optimized with the test data. 
Since these two steps are performed independently, the computational complexity can be analyzed separately. 
{
The training time (2.67 hours) for SAME in Table~\ref{tab:comparison_SOTA} refers to the time taken for the SFUDA process. 
And the training time for the pseudo-label picker has not been included, as the picker is constructed independently using public datasets and remains frozen throughout the SFUDA process.
}
Despite the moderate computational costs, the training and inference of SAME are efficient compared to the SOTA algorithms.


\subsubsection{Benefits for downstream tasks}
Boosting downstream clinical analysis and diagnosis is a key motivation for medical image enhancement. 
Therefore, vessel segmentation and fundus disease diagnosis were carried out following the enhancement to demonstrate the benefits of the proposed SAME.
Fig.~\ref{fig:comparison} (c) exhibits visualized results, and quantitative analysis is summarized in Table~\ref{tab:comparison_seg}.

\textbf{Vessel segmentation.}
Vessel segmentation is executed by a U-Net learned from DRIVE to validate the effect of enhancement on medical image analysis.
The segmentation results of the reference images are used as ground truth to quantify the segmentation performance.
DICE and IoU are computed as metrics, and those of raw low-quality images are also presented in Table~\ref{tab:comparison_seg} as a benchmark.
Fig.~\ref{fig:comparison} expresses visualized segmentation results.

Medical image enhancement highlights the distinctiveness of fundus vessels in Fig.~\ref{fig:comparison}, leading to improved segmentation results. 
Specifically, by incorporating explicit structure constraints, MAGE-Net~\cite{guo2023bridging}, SCR-Net~\cite{li2022structure}, and SAME have enjoyed advantages in improving vessel segmentation. 
Consistently, these three algorithms respectively achieve the superior quantitative segmentation results in the three datasets according to Table~\ref{tab:comparison_seg}.
Additionally, by leveraging a robust source model and SFUDA, SAME outperforms other comparative algorithms in terms of average segmentation results.

\textbf{Fundus disease diagnosis.}
Fundus-iSee was employed to investigate the diagnosis improvement achieved by SAME.
Fundus-iSee includes five subsets, i.e. normal fundus (5868 clear VS 1902 low-quality images), age-related macular degeneration (AMD) (492 VS 228), diabetic retinopathy (DR) (181 VS 89),  glaucoma (312 VS 138), and high myopia (478 VS 312).
Five thousand clear images were randomly split to learn a diagnosis model with ResNet-50, and the rest clear and low-quality images are diagnosed by the model to draw the diagnosis benchmark.
Then the low-quality images are enhanced and diagnosed again.

The diagnosis performance is quantified by F1-score and Ckappa as provided in Table~\ref{tab:comparison_seg}.
As the low quality in Fundus-iSee mainly results from cataracts~\cite{li2022annotation}, SCR-Net~\cite{li2022structure} achieves remarkable performance. 
Meanwhile, SAME achieves superior results in Ckappa, which is a more appropriate metric for evaluating multi-class classification tasks than F1-score.
Class-wise evaluation illustrates that SAME exhibits competitive performance in each class and is further improved by SFUDA.

\begin{figure}[bp]
\begin{centering}
\includegraphics[width=0.85\linewidth]{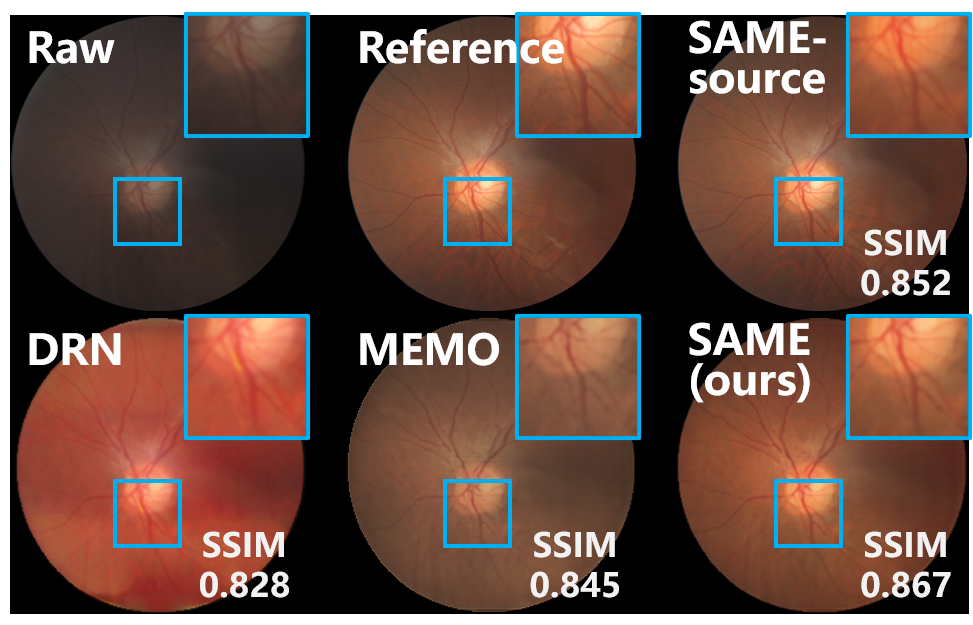}
\par\end{centering}
\caption{Visualized comparisons with SFUDA algorithms in FIQ.
}
\label{fig:SFUDA}
\end{figure}

\subsubsection{Comparisons with SFUDA benchmarks}

{SOTA algorithms for SFUDA are also compared to demonstrate the advantages of SAME.
While there have been extensive efforts to develop SFUDA paradigms for segmentation and classification tasks, the availability of algorithms available for image enhancement is still limited.
Based on the source model of SAME, two SOTA algorithms were employed as SFUDA benchmarks: DRN~\cite{yu2022source}, an SFUDA algorithm focused on image dehazing, and MEMO~\cite{zhang2022memo}, a TTA paradigm designed for robustifying against distribution shifts during test time.
The quantitative comparison of enhancement and segmentation results is summarized in Table~\ref{tab:comparison_SFUDA}, while the visual comparison is presented in Fig.~\ref{fig:SFUDA}.}

{DRN~\cite{yu2022source} and MEMO~\cite{zhang2022memo} have been implemented based on the source model of SAME.
DRN~\cite{yu2022source} leverages the frequency property and physical priors of hazy images to match the representation of real hazy domain features with that of the synthetic domain, thereby achieving SFUDA. However, since the frequency property and physical priors are primarily tailored for image dehazing, DRN~\cite{yu2022source} exhibits mediocre adaptation performance when applied to medical image enhancement.
MEMO~\cite{zhang2022memo} utilizes various data augmentations on individual data points and adapts the model by minimizing the entropy across these augmentations. In this comparison, MEMO~\cite{zhang2022memo} was introduced by augmenting the inference data, and the structure output from the source model of SAME was also utilized to incorporate the entropy minimization of MEMO~\cite{zhang2022memo}.
Despite the reasonable results in the segmentation metric of DICE, MEMO~\cite{zhang2022memo} lacks effective enhancement adaptation modules, which limits its ability to effectively address medical image enhancement tasks.
}

{In contrast to the negative adaptation observed in the benchmarks, SAME offers a robust SFUDA paradigm for medical image enhancement, enabling improvements to the source model in target sites.}

\begin{table}[tbp]
\footnotesize
\centering
\caption{Comparisons with SFUDA benchmarks on medical image enhancement and downstream segmentation}
\label{tab:comparison_SFUDA} 
\renewcommand{\arraystretch}{1}
\begin{tabular}{m{1.5cm} | m{1cm}<{\centering}| m{0.7cm}<{\centering} m{0.7cm}<{\centering} m{0.7cm}<{\centering} m{0.7cm}<{\centering}}
\hline
Algorithms & Metrics &  RF & FIQ & RCF & Avg.\\
\hline
\multirow{4}{*}{SAME-source} & \multirow{2}{*}{SSIM} & 0.750 & 0.862 & 0.761& 0.791\\
&& \raisebox{0.2em}{\scriptsize (0.047)} &  \raisebox{0.2em}{\scriptsize (0.028)} & \raisebox{0.2em}{\scriptsize (0.061)} & \raisebox{0.2em}{\scriptsize (0.050)}\\
& \multirow{2}{*}{DICE} & 0.592 & 0.619 & 0.573 & 0.595\\
&& \raisebox{0.2em}{\scriptsize (0.049)} &  \raisebox{0.2em}{\scriptsize (0.051)} & \raisebox{0.2em}{\scriptsize (0.055)} & \raisebox{0.2em}{\scriptsize (0.019)}\\
\hline

\multirow{4}{*}{DRN~\cite{yu2022source}+} & \multirow{2}{*}{SSIM} & 0.725 & 0.837 & 0.734& 0.765\\
&& \raisebox{0.2em}{\scriptsize (0.048)} &  \raisebox{0.2em}{\scriptsize (0.037)} & \raisebox{0.2em}{\scriptsize (0.067)} & \raisebox{0.2em}{\scriptsize (0.051)}\\
& \multirow{2}{*}{DICE} & 0.570 & 0.613 & 0.562 & 0.582\\
&& \raisebox{0.2em}{\scriptsize (0.053)} &  \raisebox{0.2em}{\scriptsize (0.055)} & \raisebox{0.2em}{\scriptsize (0.059)} & \raisebox{0.2em}{\scriptsize (0.022)}\\
\hline

\multirow{4}{*}{MEMO~\cite{zhang2022memo}+} & \multirow{2}{*}{SSIM} & 0.737 & 0.851 & 0.748 & 0.779\\
&& \raisebox{0.2em}{\scriptsize (0.046)} &  \raisebox{0.2em}{\scriptsize (0.027)} & \raisebox{0.2em}{\scriptsize (0.062)} & \raisebox{0.2em}{\scriptsize (0.051)}\\
& \multirow{2}{*}{DICE} &0.590 & 0.621 & 0.570& 0.594\\
&& \raisebox{0.2em}{\scriptsize (0.051)} &  \raisebox{0.2em}{\scriptsize (0.052)} & \raisebox{0.2em}{\scriptsize (0.057)} & \raisebox{0.2em}{\scriptsize (0.021)}\\
\hline

\multirow{4}{*}{SAME (ours)} & \multirow{2}{*}{SSIM} &\textbf{0.770} & \textbf{0.873} &\textbf{0.771}& \textbf{0.805}\\
&& \raisebox{0.2em}{\scriptsize (0.044)} &  \raisebox{0.2em}{\scriptsize (0.023)} & \raisebox{0.2em}{\scriptsize (0.051)} & \raisebox{0.2em}{\scriptsize (0.048)}\\
& \multirow{2}{*}{DICE} & \textbf{0.610} & \textbf{0.638} & \textbf{0.578} &\textbf{0.609 }\\
&& \raisebox{0.2em}{\scriptsize (0.048)} &  \raisebox{0.2em}{\scriptsize (0.053)} & \raisebox{0.2em}{\scriptsize (0.044)} & \raisebox{0.2em}{\scriptsize (0.025)}\\
\hline
\end{tabular}%
\end{table}

\subsection{Ablation studies}
{Comprehensive ablation studies are conducted on the setting of the source model and the adaptation process, as well as the designed modules, to thoroughly interpret the effectiveness of SAME.}

\subsubsection{Effects from source models}
\begin{figure}[tbp]
\begin{centering}
\includegraphics[width=1\linewidth]{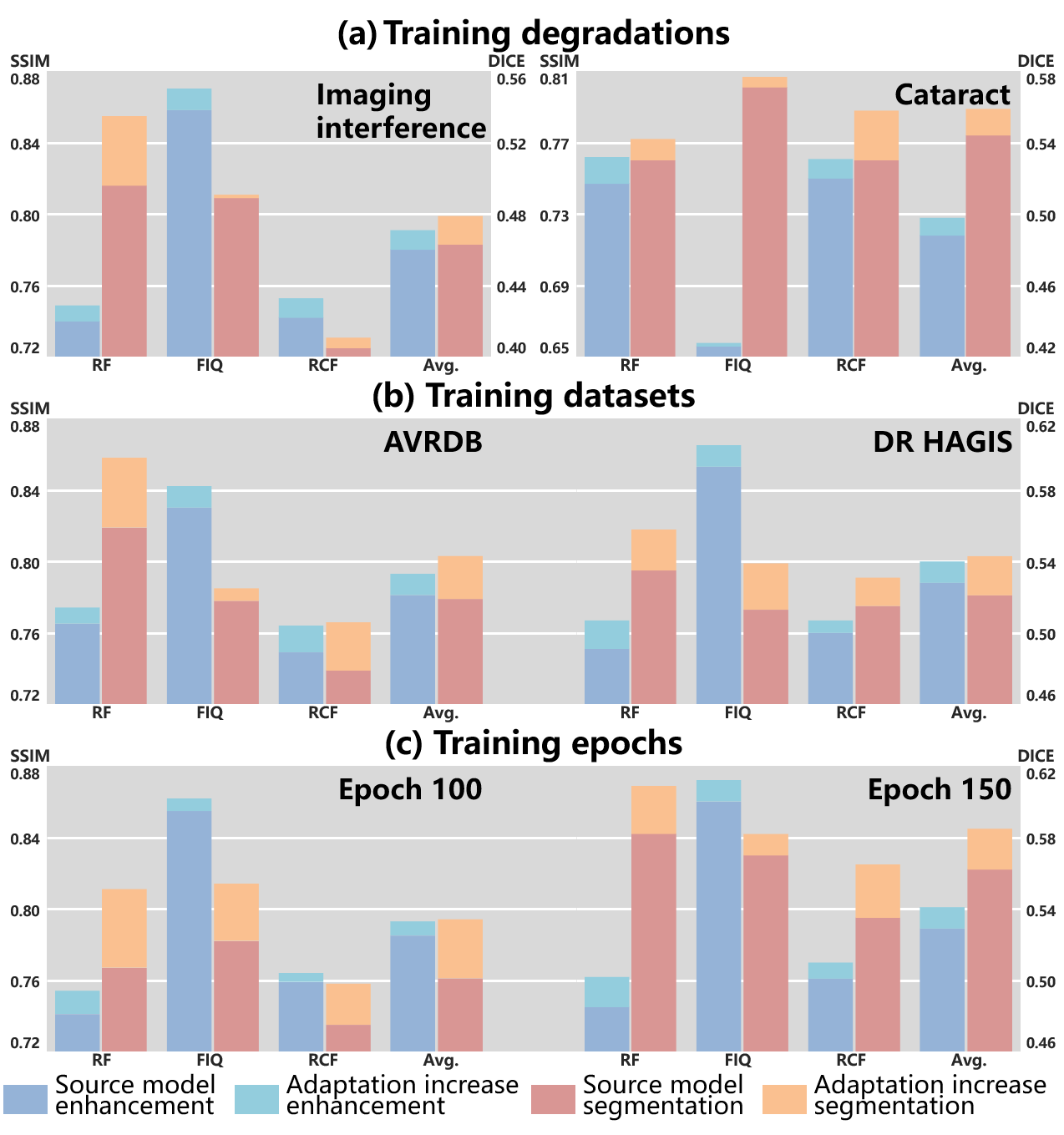}
\par\end{centering}
\caption{Enhancement (SSIM) and segmentation (DICE) performance of various source models and their corresponding adapted models. The enhancement metric SSIM is present on the left and the segmentation metric DICE is on the right.
}
\label{fig:source}
\end{figure}

To interpret the impact of the source models on SAME, the enhancement model is initialized with source models trained under various settings and then adapted using SAME. 
{The default setting for the source model of SAME involves training with synthetic data generated from DRIVE with simulated degradations, including imaging interference and cataracts, for a total of 200 epochs.}
The setting analysis comprises:
1) Degradation models of imaging interference and cataracts are independently employed to synthesize the training data for the source model.
2) The training data from DRIVE are replaced by AVRDB and DR HAGIS to train the source models.
3) Two different epoch numbers, 100 and 150, are used to train the source models.
Fig.~\ref{fig:source} summarizes the enhancement {and segmentation} performance of the various source models and the corresponding adapted models by SAME.

\textbf{Training degradations.}
{Degradations in fundus images can be categorized as imaging interference and cataracts, based on their underlying triggers. 
Furthermore, imaging interference encompasses image blur, light disturbance, and retinal artifacts. 
RF mainly involves blur and light disturbance~\cite{RFormer}.
FIQ suffers from all three types of imaging interference~\cite{liu2022degradation}.
RCF primarily encounters cataracts~\cite{li2022structure}.
As a result, the performance on the three datasets is influenced by the specific degradations used for synthesizing training data.}

{Fig.~\ref{fig:source} (a) illustrates that the degradation types have a substantial impact on the performance in FIQ, while relatively stable performance is observed in RF and RCF. 
Compared to imaging interference, only synthesizing cataracts significantly limits the performance of the source model significantly in FIQ.
The similarity between image blur and cataracts contributes to the relatively stable enhancement results in RF and RCF.
Moreover, image blur and cataracts exert a more pronounced impact on the segmentation performance compared to other types of imaging interference.
Consequently, the training data for cataract elimination demonstrates advantages in segmentation performance.}

{SAME demonstrates robust improvements in both enhancement and segmentation across all three datasets, indicating its stable progress in performance.
}

\textbf{Training datasets.}
{To assess the influence of dataset selection in the source domain on performance, DRIVE is substituted with AVRDB and DR HAGIS. To ensure a fair comparison, we synthesized 16 degraded samples randomly from each high-quality sample of DR HAGIS, while 8 samples were from that of AVRDB, taking into consideration the dataset volume. }

{As depicted in Fig.~\ref{fig:source} (b), AVRDB and DR HAGIS demonstrate comparable overall performance, albeit with some fluctuations on individual datasets. These performance fluctuations could be attributed to the homogeneity of the training and test data, which may lead to performance spikes on specific datasets. SAME demonstrates robustness to the datasets used for training the source model, resulting in consistent adaptive capabilities across various datasets. This robustness mitigates the inferior performance of source models observed in certain datasets to a certain extent.}


\textbf{Training epochs.}
{We present a comparison of the adaptation process between the source model trained for 100 and 150 epochs, respectively. Fig.~\ref{fig:source} (c) clearly demonstrates that the source model is insufficiently trained at the 100 epoch compared to the model at the 150 epoch, resulting in inferior performance, particularly in terms of segmentation.
On the other hand, the source model trained for 150 epochs is reaching convergence as it exhibits similar performance to the default setting of 200 epochs.}

{A noteworthy observation in this scenario is that the progress achieved by SAME is affected by the training epochs of the source model, which differs from the observation made regarding training degradations and datasets.
The difference in these observations may stem from the fact that degradation and datasets alter the nature of the source training data without affecting the capacity of SAME. However, the undertrained source model in the current case impacts the initialization of SAME, leading to a hindrance in the adaptation process.}

{Moreover,  it is observed that the segmentation progress is relatively less affected by the undertrained source model compared to enhancement, demonstrating a greater potential for improvement. This observation leads us to speculate that the inherent difficulty of SFUDA for enhancement surpasses that of segmentation.}

In summary, the training data plays a significant role in shaping the performance of the source model. 
Fortunately, despite training degradation and datasets impacting the adaptation of SAME for specific datasets, the overall performance remains robust across various datasets.
In contrast, the extent of training substantially affects both the initialization and the subsequent adaptation, emphasizing the importance of using a fully trained source model for the initialization of SAME.

\begin{table*}[htbp]
\footnotesize
\centering
\caption{Ablation study of the proposed modules in SAME.}
\label{tab:ablation} 
\renewcommand{\arraystretch}{0.9}
\begin{tabular}{m{0.5cm}<{\centering} m{0.5cm}<{\centering} | m{0.5cm}<{\centering} m{0.5cm}<{\centering} m{0.5cm}<{\centering} m{0.5cm}<{\centering} || m{0.9cm}<{\centering} m{0.9cm}<{\centering} m{0.9cm}<{\centering} m{0.9cm}<{\centering} |m{0.9cm}<{\centering} m{0.9cm}<{\centering} m{0.9cm}<{\centering} m{0.9cm}<{\centering}}
\hline
\multicolumn{2}{c|}{Source model} & \multicolumn{4}{c||}{Target Adaptation} & \multicolumn{4}{c|}{Enhancement (SSIM)} & \multicolumn{4}{c}{Segmentation (DICE)}\\
\hline
\rule{0pt}{1.2em} \raisebox{0.2em}
{$\mathcal{L}^S_1$} & $\mathcal{L}^S_{CE}$ & $\mathcal{L}^T_1$ & $\mathcal{L}^T_{CE}$& IQA & ISD & RF & FIQ & RCF & Avg. & RF & FIQ & RCF & Avg.\\
\hline
\multirow{2}{*}{$\surd$} &&&&&&   0.745 &   0.858 &  0.755 &  0.786 
&  0.579 &  0.612 &  0.565 & 0.585\\
&&&&&&   \raisebox{0.2em}{\scriptsize (0.049)} &  \raisebox{0.2em}{\scriptsize (0.030)} & \raisebox{0.2em}{\scriptsize (0.068)} & \raisebox{0.2em}{\scriptsize (0.051)} 
&  \raisebox{0.2em}{\scriptsize (0.061)}  & \raisebox{0.2em}{\scriptsize (0.055)} &  \raisebox{0.2em}{\scriptsize (0.060)} & \raisebox{0.2em}{\scriptsize (0.020)}\\

\multirow{2}{*}{$\surd$} && \multirow{2}{*}{$\surd$} &&&&   0.746 &   0.860  &  0.760  &  0.789
&  0.582 &  0.615 &  0.569 & 0.589\\
&&&&&&   \raisebox{0.2em}{\scriptsize (0.053)} &  \raisebox{0.2em}{\scriptsize (0.029)} & \raisebox{0.2em}{\scriptsize (0.066)} & \raisebox{0.2em}{\scriptsize (0.051)} 
&  \raisebox{0.2em}{\scriptsize (0.062)}  & \raisebox{0.2em}{\scriptsize (0.053)} &  \raisebox{0.2em}{\scriptsize (0.059)} & \raisebox{0.2em}{\scriptsize (0.019)}\\

\multirow{2}{*}{$\surd$} && \multirow{2}{*}{$\surd$} && \multirow{2}{*}{$\surd$} &&  0.751  &  0.866  & 0.764  &  0.794
&  0.592 &  0.622 &  0.572 & 0.595\\
&&&&&&   \raisebox{0.2em}{\scriptsize (0.050)} &  \raisebox{0.2em}{\scriptsize (0.027)} & \raisebox{0.2em}{\scriptsize (0.060)} & \raisebox{0.2em}{\scriptsize (0.051)} 
&  \raisebox{0.2em}{\scriptsize (0.058)}  & \raisebox{0.2em}{\scriptsize (0.051)} &  \raisebox{0.2em}{\scriptsize (0.058)} & \raisebox{0.2em}{\scriptsize (0.021)}\\
\hline

\multirow{2}{*}{$\surd$} & \multirow{2}{*}{$\surd$} &&&&&  0.750  &  0.862  &  0.761 &  0.791  
&  0.592  &  0.619  &  0.573  &  0.595 \\
&&&&&&   \raisebox{0.2em}{\scriptsize (0.047)} &  \raisebox{0.2em}{\scriptsize (0.028)} & \raisebox{0.2em}{\scriptsize (0.061)} & \raisebox{0.2em}{\scriptsize (0.050)} 
&  \raisebox{0.2em}{\scriptsize (0.049)}  & \raisebox{0.2em}{\scriptsize (0.051)} &  \raisebox{0.2em}{\scriptsize (0.055)} & \raisebox{0.2em}{\scriptsize (0.019)}\\

\multirow{2}{*}{$\surd$} &  \multirow{2}{*}{$\surd$} & \multirow{2}{*}{$\surd$} &&&&  0.746  &  0.865  &  0.765  &  0.792 
&  0.586 &  0.620 &  0.573 & 0.593\\
&&&&&&   \raisebox{0.2em}{\scriptsize (0.049)} &  \raisebox{0.2em}{\scriptsize (0.029)} & \raisebox{0.2em}{\scriptsize (0.058)} & \raisebox{0.2em}{\scriptsize (0.052)} 
&  \raisebox{0.2em}{\scriptsize (0.048)}  & \raisebox{0.2em}{\scriptsize (0.053)} &  \raisebox{0.2em}{\scriptsize (0.055)} & \raisebox{0.2em}{\scriptsize (0.020)}\\

\multirow{2}{*}{$\surd$} & \multirow{2}{*}{$\surd$} & \multirow{2}{*}{$\surd$} && \multirow{2}{*}{$\surd$} &&  0.754  &  0.870  & 0.766 & 0.797  
&  0.595 &  0.625 &  0.574  & 0.598\\
&&&&&&   \raisebox{0.2em}{\scriptsize (0.051)} &  \raisebox{0.2em}{\scriptsize (0.027)} & \raisebox{0.2em}{\scriptsize (0.055)} & \raisebox{0.2em}{\scriptsize (0.052)} 
&  \raisebox{0.2em}{\scriptsize (0.045)}  & \raisebox{0.2em}{\scriptsize (0.051)} &  \raisebox{0.2em}{\scriptsize (0.055)} & \raisebox{0.2em}{\scriptsize (0.021)}\\

\multirow{2}{*}{$\surd$} &  \multirow{2}{*}{$\surd$} & \multirow{2}{*}{$\surd$} & \multirow{2}{*}{$\surd$} &&&  0.753  &  0.867  &  0.766  &  0.795 
& 0.595 &  0.626 &  0.575 & 0.599 \\
&&&&&&   \raisebox{0.2em}{\scriptsize (0.051)} &  \raisebox{0.2em}{\scriptsize (0.033)} & \raisebox{0.2em}{\scriptsize (0.057)} & \raisebox{0.2em}{\scriptsize (0.056)} 
&  \raisebox{0.2em}{\scriptsize (0.046)}  & \raisebox{0.2em}{\scriptsize (0.052)} &  \raisebox{0.2em}{\scriptsize (0.056)} & \raisebox{0.2em}{\scriptsize (0.025)}\\

\multirow{2}{*}{$\surd$} &  \multirow{2}{*}{$\surd$} & \multirow{2}{*}{$\surd$} & \multirow{2}{*}{$\surd$} & \multirow{2}{*}{$\surd$} &&  0.767  &  0.871  &  0.769  &  0.802 
&    0.596 &  0.630 &  0.575 & 0.600\\
&&&&&&   \raisebox{0.2em}{\scriptsize (0.046)} &  \raisebox{0.2em}{\scriptsize (0.025)} & \raisebox{0.2em}{\scriptsize (0.053)} & \raisebox{0.2em}{\scriptsize (0.049)} 
&  \raisebox{0.2em}{\scriptsize (0.047)}  & \raisebox{0.2em}{\scriptsize (0.048)} &  \raisebox{0.2em}{\scriptsize (0.051)} & \raisebox{0.2em}{\scriptsize (0.023)}\\

\multirow{2}{*}{$\surd$} &  \multirow{2}{*}{$\surd$} & \multirow{2}{*}{$\surd$} & \multirow{2}{*}{$\surd$} && \multirow{2}{*}{$\surd$} &   0.755  &  0.872   &  0.767  & 0.798
&    0.605 &  0.635 &  0.576 & 0.605\\
&&&&&&   \raisebox{0.2em}{\scriptsize (0.055)} &  \raisebox{0.2em}{\scriptsize (0.024)} & \raisebox{0.2em}{\scriptsize (0.055)} & \raisebox{0.2em}{\scriptsize (0.053)} 
&  \raisebox{0.2em}{\scriptsize (0.049)}  & \raisebox{0.2em}{\scriptsize (0.056)} &  \raisebox{0.2em}{\scriptsize (0.047)} & \raisebox{0.2em}{\scriptsize (0.024)}\\

\multirow{2}{*}{$\surd$} &  \multirow{2}{*}{$\surd$} & \multirow{2}{*}{$\surd$} & \multirow{2}{*}{$\surd$} & \multirow{2}{*}{$\surd$} & \multirow{2}{*}{$\surd$} &  0.770  &  0.873  & 0.771 &  0.805  &  0.610  &  0.638  &  0.578  & 0.609 \\
&&&&&& \raisebox{0.2em}{\scriptsize (0.044)} &  \raisebox{0.2em}{\scriptsize (0.023)} & \raisebox{0.2em}{\scriptsize (0.051)} & \raisebox{0.2em}{\scriptsize (0.048)}  
&  \raisebox{0.2em}{\scriptsize (0.048)}  & \raisebox{0.2em}{\scriptsize (0.053)} &  \raisebox{0.2em}{\scriptsize (0.044)} & \raisebox{0.2em}{\scriptsize (0.025)}\\
\hline
\end{tabular}%
\end{table*}

\subsubsection{Adaptation epochs and samples}
\begin{figure}[tbp]
\begin{centering}
\includegraphics[width=1\linewidth]{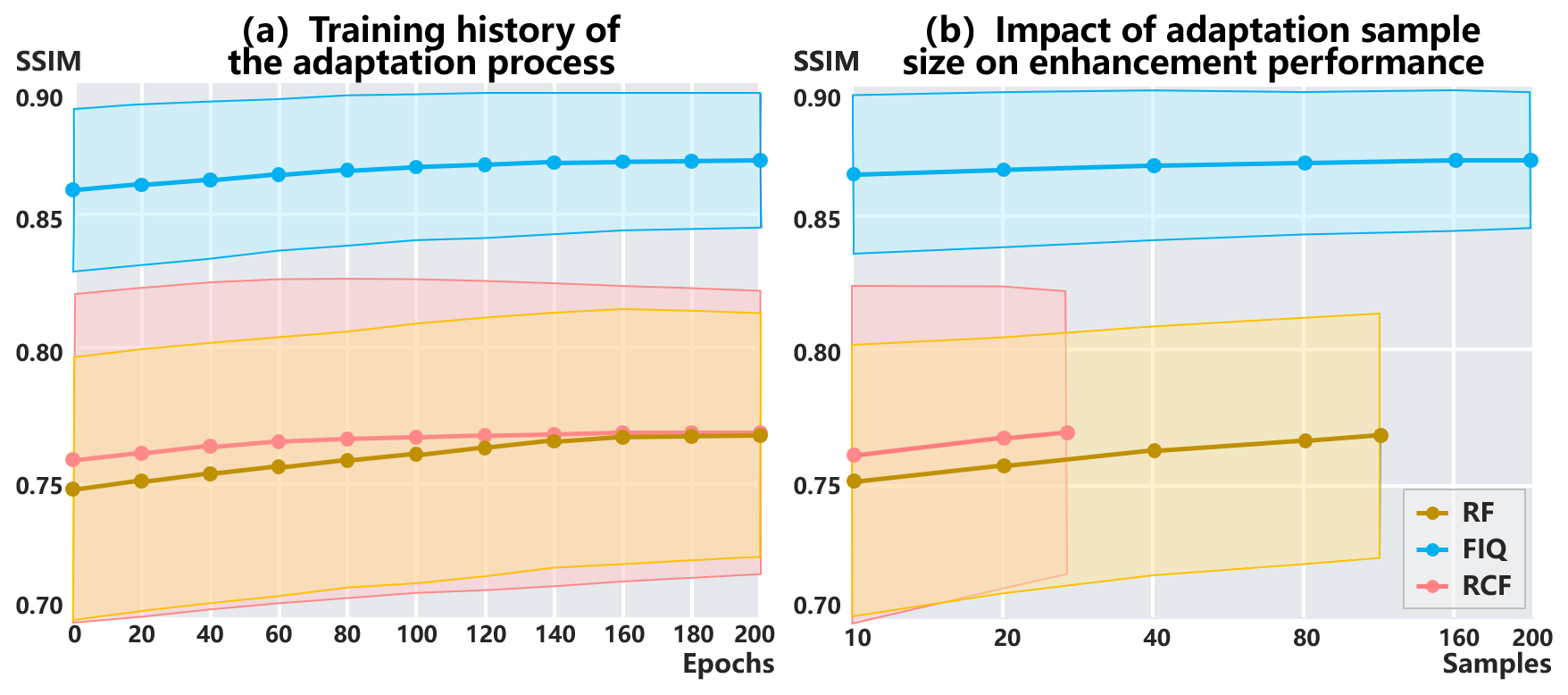}
\par\end{centering}
\caption{The training history of the adaptation process, and the performance of enhancement with adaptation sample size. 
}
\label{fig:samples}
\end{figure}

{To gain insights into the adaptation process, we investigate the performance affected by adaptation epochs and samples. The training history of the adaptation process and the performance variance with the adaptation samples are visualized in Fig.~\ref{fig:samples}.}

\textbf{Adaptation epochs.}
{The adaptation process is performed by referencing the configuration of the source model training. 
This involves utilizing the Adam optimizer without early stopping. 
The learning rate is 0.001 in the first 150 epochs, and gradually decreases to 0 during the final 50 epochs.}

{Fig.~\ref{fig:samples} (a) exhibits the history of the adaptation process.
SAME achieves convergence within 200 adaptation epochs across all datasets. Convergence is reached earlier in RF and FIQ compared to RCF, likely due to the difference in sample sizes.}

\textbf{Adaptation samples.} 
{Fig.~\ref{fig:samples} (b) provides a visual representation of the enhancement performance and variance influenced by the adaptation sample size.
It is observed that as the sample size increases, the performance exhibits positive progress. Optimal performance is achieved when utilizing all available samples for adaptation.}

{Furthermore, performance convergence is observed at the sample size of 160 in FIQ, which can be attributed to its larger data volume compared to the other two datasets. 
This insight highlights the importance of using a representative sample size during the adaptation process, as it facilitates the generalization of adapted models.}

\subsubsection{Module Ablation}
Ablation studies against the modules in SAME are summarized in Table~\ref{tab:ablation} to verify their effectiveness.
{The source model with and without the segmentation decoder are respectively employed as the initialization.
Subsequently, the adaptation modules are individually implemented to validate their effectiveness.}

{The source model without the segmentation decoder is exclusively trained using $\mathcal{L}^S_1$ on the source domain. For adaptation, distillation with $\mathcal{L}^T_1$ is performed on target domains, and pseudo-labels can be selected using the IQA.
The absence of the segmentation decoder hinders the source model from effectively preserving structures, impacting both the enhancement and segmentation. The utilization of the IQA for pseudo-label selection further boosts the adaptation performance.}

{By considering both image enhancement and structure preservation using $\mathcal{L}^S_1$ and $\mathcal{L}^S_{CE}$, the capacity of the source model is promoted. However, ignoring structure preservation during adaptation (e.g., solely utilizing $\mathcal{L}^T_1$) may lead to negative adaptation, as observed in RF. The collaborative use of $\mathcal{L}^T_1$ and $\mathcal{L}^T_{CE}$ ensures positive adaptation on target domains. Furthermore, the selection of pseudo-labels with the IQA and the ISD contributes to advancements in adaptation. The IQA demonstrates advantages in boosting image enhancement, while the ISD reasonably promotes structure preservation.}

\begin{figure*}[htbp]
\begin{centering}
\includegraphics[width=0.9\linewidth]{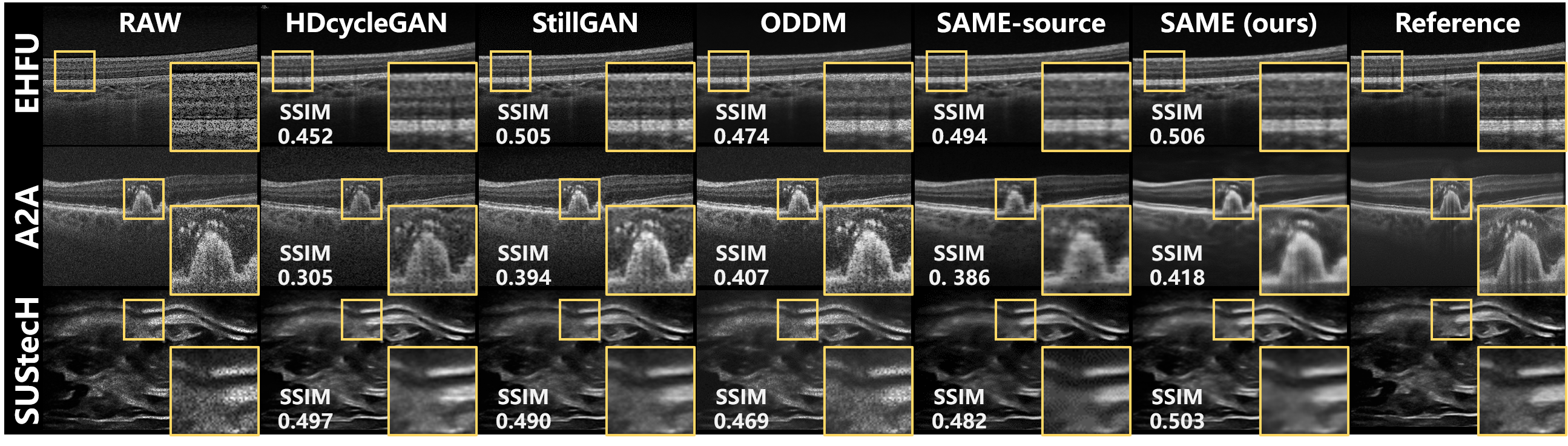}
\par\end{centering}
\caption{Visualized comparison of enhancement results of OCT and ultrasound images.
The top two rows exhibit the enhanced images of OCT from EHFU and A2A, while the bottom row displays the enhancement results of ultrasound images from SUStecH.
{The enhancement progress achieved by SAME indicates its versatility across diverse medical imaging modalities.}
}
\label{fig:comparison_oct}
\end{figure*}

\subsection{Versatility for various modalities}
The versatility of SAME is demonstrated through the enhancement of various medical imaging modalities, and quantitative and visualized comparisons with the enhancement algorithms for the modalities are provided in Table~\ref{tab:comparison_oct} and Fig.~\ref{fig:comparison_oct}.

Considering that both OCT and ultrasound images are affected by speckle noise~\cite{bioucas2010multiplicative}, we compare the same enhancement algorithms for both modalities. 
{As indicated in Table~\ref{tab:datasets}, the clear samples from EHFU and SUStecH are collected and collaborated with the degradation model described in \cite{bioucas2010multiplicative} to synthesize paired training data.
As a result, the unpaired data-based algorithms are free from domain shifts on EHFU and SUStecH, since their training and test data are all from these two datasets. 
And the synthetic data-based algorithms also suffer fewer domain shifts on EHFU and SUStecH compared to A2A.
}

\begin{table}[tbp]
\footnotesize
\centering
\caption{Comparisons on OCT and ultrasound enhancement}
\label{tab:comparison_oct} 
\renewcommand{\arraystretch}{1}
\begin{tabular}{p{2.2cm} | p{0.8cm}<{\centering} p{0.8cm}<{\centering} p{0.8cm}<{\centering} p{0.8cm}<{\centering}} 
\hline
\multirow{2}{*}{Algorithms} & \multicolumn{4}{c}{Enhancement (SSIM)}\\ 
\cline{2-5}
& EHFU & A2A & SUStecH & Avg. \\
\hline
\multirow{2}{*}{HDcycleGAN~\cite{manakov2019noise}} &  0.435  & 0.308 & \textbf{0.505} & 0.416\\
& \raisebox{0.2em}{\scriptsize (0.057)} &  \raisebox{0.2em}{\scriptsize (0.062)} & \raisebox{0.2em}{\scriptsize (0.033)} & \raisebox{0.2em}{\scriptsize (0.082)}\\
\multirow{2}{*}{StillGAN~\cite{ma2021structure}}  &\textbf{0.507} & 0.388 & 0.484 & 0.460 \\
& \raisebox{0.2em}{\scriptsize (0.050)} &  \raisebox{0.2em}{\scriptsize (0.055)} & \raisebox{0.2em}{\scriptsize (0.038)} & \raisebox{0.2em}{\scriptsize (0.052)}\\
\multirow{2}{*}{ODDM~\cite{hu2022unsupervised}} & 0.476 & 0.406 & 0.461 & 0.448\\
& \raisebox{0.2em}{\scriptsize (0.055)} &  \raisebox{0.2em}{\scriptsize (0.049)} & \raisebox{0.2em}{\scriptsize (0.037)} & \raisebox{0.2em}{\scriptsize (0.030)}\\
\hline
\multirow{2}{*}{SAME-source} & 0.491 & 0.378 & 0.479 & 0.449\\
& \raisebox{0.2em}{\scriptsize (0.053)} &  \raisebox{0.2em}{\scriptsize (0.052)} & \raisebox{0.2em}{\scriptsize (0.039)} & \raisebox{0.2em}{\scriptsize (0.051)}\\
\multirow{2}{*}{SAME (ours)} & 0.495 & \textbf{0.408}& 0.490 & \textbf{0.464} \\
& \raisebox{0.2em}{\scriptsize (0.056)} &  \raisebox{0.2em}{\scriptsize (0.051)} & \raisebox{0.2em}{\scriptsize (0.038)} & \raisebox{0.2em}{\scriptsize (0.040)}\\
\hline
\end{tabular}%
\end{table}

{
Therefore, it is no surprise that the enhancement performance on EHFU and SUStecH would generally surpass that on A2A, given the effect of domain shifts.
Furthermore, the absence of domain shifts enables HDcycleGAN~\cite{manakov2019noise} and StillGAN~\cite{ma2021structure} to deliver impressive results on EHFU and SUStecH, as summarized in Table~\ref{tab:comparison_oct}.
The source model provides a reasonable initialization for SAME across all datasets, and further progress is achieved by SAME (visualized in Fig.~\ref{fig:comparison_oct}), particularly on A2A, through the additional adaptation of the model to each specific dataset using SFUDA.
These findings indicate that SAME exhibits versatility across various medical imaging modalities.
}

\section{Discussions}
Enhancement algorithms based on unpaired and synthetic training data have been developed to boost clinical observation and diagnosis. 
However, deploying these algorithms in the clinic poses a challenge due to domain shifts.
Independent identically distributed training and test data without domain shifts are impractical in the clinical setting, and incorporating clinical test data to compress domain shifts leads to the high costs and privacy concerns associated with data collection.
{To alleviate these challenges, we propose an SFUDA paradigm for medical image enhancement, named SAME. SAME leverages a source model trained on synthetic data to initialize a teacher-student model, which implements SFUDA on target sites through knowledge distillation. By avoiding the centralized storage and training of clinical data, SAME reduces the need for data storage and transmission, thereby mitigating privacy risks.
Although some privacy issues remain, such as the risk associated with the release of source models, addressing risks related to clinical data usage is still significant. In our future studies, we will strive to further minimize the risk from source models.}

{To implement SAME, a decent source model is first learned from synthetic training data for initialization, and then the designed knowledge distillation framework and pseudo-label picker collaborate to conduct SFUDA for medical image enhancement.}
In the experiment, comparisons with SOTA algorithms were executed not only on the enhancement task but also on the downstream segmentation and diagnosis tasks to validate the benefits of SAME.
The results confirm that SAME significantly improved the enhancement model at the targeted sites during the inference phase, without compromising data collection and privacy protection. 
Moreover, SAME also provided a notable boost to segmentation and diagnosis tasks.

{Despite the superior performance of SAME, existing algorithms also achieved remarkable outstanding results on certain datasets due to their specifically designed modules.
However, these specific modules may not be easily transferable across various image datasets and modalities.}
In contrast, SAME offers a versatile paradigm for enhancing medical images, optimizing the enhancement model during the inference phase. Its versatility was validated by applying SAME to three distinct medical image modalities: fundus photography, OCT, and ultrasound images. Remarkably, SAME showcased superior enhancement performance across all three modalities.

{Comprehensive ablation studies were also conducted on the configuration of the source model and the adaptation process, as well as the designed modules, to thoroughly investigate the effectiveness and potential of SAME.
Training data has proved to have notable effects on the source model and the initialization of SAME.
In spite of the adaptation variations on specific datasets caused by training degradation and datasets, the overall performance of SAME  remains robust across various datasets.
Notably, the extent of training substantially affects both the initialization and the subsequent adaptation of SAME, highlighting the necessity of utilizing a fully trained source model for the initialization.
During the adaptation phase, the sample size has been verified to have effects on the convergence speed and the performance of the model after adaptation.
Regarding the designed modules, it has been confirmed that the segmentation decoder imposes structure-preserving in the source model. 
Then the proposed knowledge distillation paradigm effectively adapts the teacher-student model to the target sites.
Additionally, the pseudo-labels selected by the IQA and the ISD ensure a positive adaptation by SAME across datasets.}

\begin{figure}[htbp]
\begin{centering}
\includegraphics[width=1\linewidth]{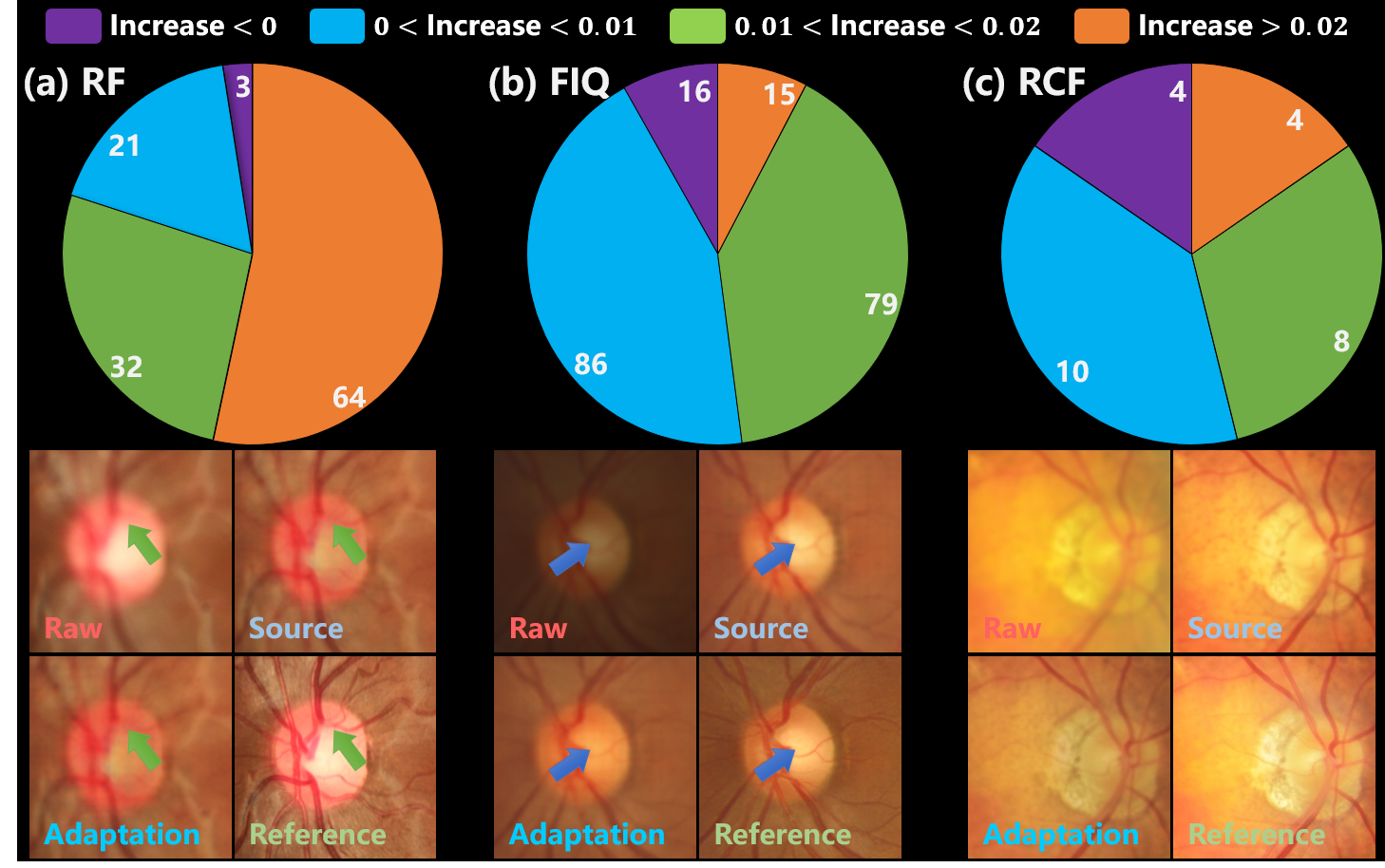}
\par\end{centering}
\caption{Sample-wise adaptation effects and failure cases. 
The pie chart counts the number of adapted samples based on their SSIM increase. The purple portion represents the failure cases and the representation ones are displayed.
}
\label{fig:limit}
\end{figure}

On the other hand, Fig.~\ref{fig:limit} illustrates the sample-wise adaptation effects, where the purple portion of the pie chart summarizes the number of failure cases.
As exhibited by the cases in Fig.~\ref{fig:limit}, the adaptation is ineffective for excessively blurred vessels in RF. Furthermore, some negative adaptation is observed in the contrast of FIQ samples and the brightness of RCF ones. These negative adapted cases are believed to be attributed to intra-dataset inconsistency among samples, such as variations in contrast and exposure. Thus the inconsistency can potentially mislead the knowledge distillation process to negative adaptation.
For the three datasets, the number, rate, and SSIM decline boundary of negative adapted samples are as follows: (3, 2.5\%, -0.001), (16, 8.2\%, -0.019), and (4, 15.4\%, -0.035) respectively.

Additionally, while SAME introduces an SFUDA paradigm for medical image enhancement, allowing for adaptive enhancement models during the inference phase, the adaptive model training with target data remains essential.
In future work, we will delve into test-time adaptation paradigms to enable the generalization of models across various target domains without explicit training with target data.


\section{Conclusions}
Medical images are often subject to quality degradation, negatively impacting clinical observation and diagnosis. 
Though enhancement algorithms have been proposed, they always require well pre-training before deployment, while failing to capitalize on inference data and promise performance on unseen data.
To address these challenges, we proposed SAME, which utilizes SFUDA to adapt and optimize enhancement models using test data in the inference phase. 
Extensive experiments were executed to interpret the advantages and effectiveness of SAME. By implementing SFUDA, SAME achieved superior performance on both the enhancement task and downstream tasks, without the additional burden of data collection and privacy protections.

\bibliographystyle{IEEEtran}
\bibliography{refer}

\begin{thebibliography}{10}
\providecommand{\url}[1]{#1}
\csname url@samestyle\endcsname
\providecommand{\newblock}{\relax}
\providecommand{\bibinfo}[2]{#2}
\providecommand{\BIBentrySTDinterwordspacing}{\spaceskip=0pt\relax}
\providecommand{\BIBentryALTinterwordstretchfactor}{4}
\providecommand{\BIBentryALTinterwordspacing}{\spaceskip=\fontdimen2\font plus
\BIBentryALTinterwordstretchfactor\fontdimen3\font minus \fontdimen4\font\relax}
\providecommand{\BIBforeignlanguage}[2]{{%
\expandafter\ifx\csname l@#1\endcsname\relax
\typeout{** WARNING: IEEEtran.bst: No hyphenation pattern has been}%
\typeout{** loaded for the language `#1'. Using the pattern for}%
\typeout{** the default language instead.}%
\else
\language=\csname l@#1\endcsname
\fi
#2}}
\providecommand{\BIBdecl}{\relax}
\BIBdecl

\bibitem{li2021applications}
T.~Li, W.~Bo, C.~Hu, H.~Kang, H.~Liu, K.~Wang, and H.~Fu, ``Applications of deep learning in fundus images: A review,'' \emph{Medical Image Analysis}, p. 101971, 2021.

\bibitem{conzelmann2022comparison}
J.~Conzelmann, U.~Genske, A.~Emig, M.~Scheel, B.~Hamm, and P.~Jahnke, ``Comparison of low-contrast detectability between uniform and anatomically realistic phantoms—influences on ct image quality assessment,'' \emph{European Radiology}, vol.~32, no.~2, pp. 1267--1275, 2022.

\bibitem{liu2022understanding}
H.~Liu, H.~Li, X.~Wang, H.~Li, M.~Ou, L.~Hao, Y.~Hu, and J.~Liu, ``Understanding how fundus image quality degradation affects cnn-based diagnosis,'' in \emph{2022 44th Annual International Conference of the IEEE Engineering in Medicine \& Biology Society (EMBC)}.\hskip 1em plus 0.5em minus 0.4em\relax IEEE, 2022, pp. 438--442.

\bibitem{welikala2017automated}
R.~Welikala, M.~Fraz, M.~Habib, S.~Daniel-Tong, M.~Yates, P.~Foster, P.~Whincup, A.~R. Rudnicka, C.~G. Owen, D.~Strachan \emph{et~al.}, ``Automated quantification of retinal vessel morphometry in the uk biobank cohort,'' in \emph{2017 Seventh International Conference on Image Processing Theory, Tools and Applications (IPTA)}.\hskip 1em plus 0.5em minus 0.4em\relax IEEE, 2017, pp. 1--6.

\bibitem{mitra2018enhancement}
A.~Mitra, S.~Roy, S.~Roy, and S.~K. Setua, ``Enhancement and restoration of non-uniform illuminated fundus image of retina obtained through thin layer of cataract,'' \emph{Computer methods and programs in biomedicine}, vol. 156, pp. 169--178, 2018.

\bibitem{RFormer}
Z.~Deng, Y.~Cai, L.~Chen, Z.~Gong, Q.~Bao, X.~Yao, D.~Fang, W.~Yang, S.~Zhang, and L.~Ma, ``Rformer: Transformer-based generative adversarial network for real fundus image restoration on a new clinical benchmark,'' \emph{IEEE Journal of Biomedical and Health Informatics}, vol.~26, no.~9, pp. 4645--4655, 2022.

\bibitem{ma2021structure}
Y.~Ma, J.~Liu, Y.~Liu, H.~Fu, Y.~Hu, J.~Cheng, H.~Qi, Y.~Wu, J.~Zhang, and Y.~Zhao, ``Structure and illumination constrained gan for medical image enhancement,'' \emph{IEEE Transactions on Medical Imaging}, 2021.

\bibitem{cheng2021secret}
P.~Cheng, L.~Lin, Y.~Huang, J.~Lyu, and X.~Tang, ``I-secret: Importance-guided fundus image enhancement via semi-supervised contrastive constraining,'' in \emph{International Conference on Medical Image Computing and Computer-Assisted Intervention}.\hskip 1em plus 0.5em minus 0.4em\relax Springer, 2021, pp. 87--96.

\bibitem{li2022annotation}
H.~Li, H.~Liu, Y.~Hu, H.~Fu, Y.~Zhao, H.~Miao, and J.~Liu, ``An annotation-free restoration network for cataractous fundus images,'' \emph{IEEE Transactions on Medical Imaging}, 2022.

\bibitem{shen2020modeling}
Z.~Shen, H.~Fu, J.~Shen, and L.~Shao, ``Modeling and enhancing low-quality retinal fundus images,'' \emph{IEEE transactions on medical imaging}, vol.~40, no.~3, pp. 996--1006, 2020.

\bibitem{li2023generic}
H.~Li, H.~Liu, H.~Fu, Y.~Xu, H.~Shu, K.~Niu, Y.~Hu, and J.~Liu, ``A generic fundus image enhancement network boosted by frequency self-supervised representation learning,'' \emph{Medical Image Analysis}, vol.~90, p. 102945, 2023.

\bibitem{zuiderveld1994contrast}
K.~Zuiderveld, ``Contrast limited adaptive histogram equalization,'' \emph{Graphics gems}, pp. 474--485, 1994.

\bibitem{he2012guided}
K.~He, J.~Sun, and X.~Tang, ``Guided image filtering,'' \emph{IEEE transactions on pattern analysis and machine intelligence}, vol.~35, no.~6, pp. 1397--1409, 2012.

\bibitem{cheng2018structure}
J.~Cheng \emph{et~al.}, ``Structure-preserving guided retinal image filtering and its application for optic disk analysis,'' \emph{IEEE transactions on medical imaging}, vol.~37, no.~11, pp. 2536--2546, 2018.

\bibitem{zhu2017unpaired}
J.-Y. Zhu \emph{et~al.}, ``Unpaired image-to-image translation using cycle-consistent adversarial networks,'' in \emph{Proceedings of the IEEE international conference on computer vision}, 2017, pp. 2223--2232.

\bibitem{manakov2019noise}
I.~Manakov, M.~Rohm, C.~Kern, B.~Schworm, K.~Kortuem, and V.~Tresp, ``Noise as domain shift: Denoising medical images by unpaired image translation,'' in \emph{MICCAI Workshop, DART 2019, Shenzhen, China, October 13 and 17, 2019, Proceedings 1}.\hskip 1em plus 0.5em minus 0.4em\relax Springer, 2019, pp. 3--10.

\bibitem{park2020contrastive}
T.~Park, A.~A. Efros, R.~Zhang, and J.-Y. Zhu, ``Contrastive learning for unpaired image-to-image translation,'' in \emph{European Conference on Computer Vision}.\hskip 1em plus 0.5em minus 0.4em\relax Springer, 2020, pp. 319--345.

\bibitem{hu2022unsupervised}
D.~Hu, Y.~K. Tao, and I.~Oguz, ``Unsupervised denoising of retinal oct with diffusion probabilistic model,'' in \emph{Medical Imaging 2022: Image Processing}, vol. 12032.\hskip 1em plus 0.5em minus 0.4em\relax SPIE, 2022, pp. 25--34.

\bibitem{guo2023bridging}
E.~Guo, H.~Fu, L.~Zhou, and D.~Xu, ``Bridging synthetic and real images: a transferable and multiple consistency aided fundus image enhancement framework,'' \emph{IEEE Transactions on Medical Imaging}, 2023.

\bibitem{li2022structure}
H.~Li, H.~Liu, H.~Fu, H.~Shu, Y.~Zhao, X.~Luo, Y.~Hu, and J.~Liu, ``Structure-consistent restoration network for cataract fundus image enhancement,'' in \emph{MICCAI}.\hskip 1em plus 0.5em minus 0.4em\relax Springer, 2022, pp. 487--496.

\bibitem{liu2022degradation}
H.~Liu, H.~Li, H.~Fu, R.~Xiao, Y.~Gao, Y.~Hu, and J.~Liu, ``Degradation-invariant enhancement of fundus images via pyramid constraint network,'' in \emph{MICCAI}.\hskip 1em plus 0.5em minus 0.4em\relax Springer, 2022, pp. 507--516.

\bibitem{zhou2022domain}
K.~Zhou, Z.~Liu, Y.~Qiao, T.~Xiang, and C.~C. Loy, ``Domain generalization: A survey,'' \emph{IEEE Transactions on Pattern Analysis and Machine Intelligence}, 2022.

\bibitem{li2023frequency}
H.~Li, H.~Li, W.~Zhao, H.~Fu, X.~Su, Y.~Hu, and J.~Liu, ``Frequency-mixed single-source domain generalization for medical image segmentation,'' in \emph{MICCAI}.\hskip 1em plus 0.5em minus 0.4em\relax Springer, 2023, pp. 127--136.

\bibitem{guan2021domain}
H.~Guan and M.~Liu, ``Domain adaptation for medical image analysis: a survey,'' \emph{IEEE Transactions on Biomedical Engineering}, vol.~69, no.~3, pp. 1173--1185, 2021.

\bibitem{liang2023comprehensive}
J.~Liang, R.~He, and T.~Tan, ``A comprehensive survey on test-time adaptation under distribution shifts,'' \emph{arXiv preprint arXiv:2303.15361}, 2023.

\bibitem{fang2022source}
Y.~Fang, P.-T. Yap, W.~Lin, H.~Zhu, and M.~Liu, ``Source-free unsupervised domain adaptation: A survey,'' \emph{arXiv preprint arXiv:2301.00265}, 2022.

\bibitem{liang2020we}
J.~Liang, D.~Hu, and J.~Feng, ``Do we really need to access the source data? source hypothesis transfer for unsupervised domain adaptation,'' in \emph{International conference on machine learning}.\hskip 1em plus 0.5em minus 0.4em\relax PMLR, 2020, pp. 6028--6039.

\bibitem{chen2022self}
W.~Chen, L.~Lin, S.~Yang, D.~Xie, S.~Pu, and Y.~Zhuang, ``Self-supervised noisy label learning for source-free unsupervised domain adaptation,'' in \emph{2022 IEEE/RSJ International Conference on Intelligent Robots and Systems (IROS)}.\hskip 1em plus 0.5em minus 0.4em\relax IEEE, 2022, pp. 10\,185--10\,192.

\bibitem{shin2022mm}
I.~Shin, Y.-H. Tsai, B.~Zhuang, S.~Schulter, B.~Liu, S.~Garg, I.~S. Kweon, and K.-J. Yoon, ``Mm-tta: multi-modal test-time adaptation for 3d semantic segmentation,'' in \emph{Proceedings of the IEEE/CVF Conference on Computer Vision and Pattern Recognition}, 2022, pp. 16\,928--16\,937.

\bibitem{zhao2023source}
C.~Zhao, R.~Peng, and D.~Wu, ``Source-free domain adaptation (sfda) for privacy-preserving seizure subtype classification,'' \emph{IEEE Transactions on Neural Systems and Rehabilitation Engineering}, 2023.

\bibitem{liu2022source}
X.~Liu and Y.~Yuan, ``A source-free domain adaptive polyp detection framework with style diversification flow,'' \emph{IEEE Transactions on Medical Imaging}, vol.~41, no.~7, pp. 1897--1908, 2022.

\bibitem{wang2020tent}
D.~Wang, E.~Shelhamer, S.~Liu, B.~Olshausen, and T.~Darrell, ``Tent: Fully test-time adaptation by entropy minimization,'' in \emph{International Conference on Learning Representations}, 2020.

\bibitem{zhang2022memo}
M.~Zhang, S.~Levine, and C.~Finn, ``Memo: Test time robustness via adaptation and augmentation,'' \emph{Advances in Neural Information Processing Systems}, vol.~35, pp. 38\,629--38\,642, 2022.

\bibitem{yu2022source}
H.~Yu, J.~Huang, Y.~Liu, Q.~Zhu, M.~Zhou, and F.~Zhao, ``Source-free domain adaptation for real-world image dehazing,'' in \emph{Proceedings of the 30th ACM International Conference on Multimedia}, 2022, pp. 6645--6654.

\bibitem{ronneberger2015u}
O.~Ronneberger, P.~Fischer, and T.~Brox, ``U-net: Convolutional networks for biomedical image segmentation,'' in \emph{Medical Image Computing and Computer-Assisted Intervention--MICCAI 2015: 18th International Conference, Munich, Germany, October 5-9, 2015, Proceedings, Part III 18}.\hskip 1em plus 0.5em minus 0.4em\relax Springer, 2015, pp. 234--241.

\bibitem{luo2020shape}
B.~Luo, J.~Shen, S.~Cheng, Y.~Wang, and M.~Pantic, ``Shape constrained network for eye segmentation in the wild,'' in \emph{Proceedings of the IEEE/CVF Winter Conference on Applications of Computer Vision}, 2020, pp. 1952--1960.

\bibitem{fu2019evaluation}
H.~Fu, B.~Wang, J.~Shen, S.~Cui, Y.~Xu, J.~Liu, and L.~Shao, ``Evaluation of retinal image quality assessment networks in different color-spaces,'' in \emph{Medical Image Computing and Computer Assisted Intervention--MICCAI 2019: 22nd International Conference, Shenzhen, China, October 13--17, 2019, Proceedings, Part I 22}.\hskip 1em plus 0.5em minus 0.4em\relax Springer, 2019, pp. 48--56.

\bibitem{bioucas2010multiplicative}
J.~M. Bioucas-Dias and M.~A. Figueiredo, ``Multiplicative noise removal using variable splitting and constrained optimization,'' \emph{IEEE Transactions on Image Processing}, vol.~19, no.~7, pp. 1720--1730, 2010.

\end{thebibliography}

\end{document}